# Substrate-Mediated Persistent Photodoping in $WSe_2$/hBN Field-Effect Transistors Enabled by Defect States in $SiO_2$

*Sung-Ha Kim[1,†], Seong-Yeon Lee[1,†], Tae-Jeong Kim[1], Beomkyu Shin[1], Young-Jun Yu[1], Kenji Watanabe[2], Takashi Taniguchi[3], Soungmin Bae[4,*], and Ki-Ju Yee[1,*]*

*[1]Department of Physics, Chungnam National University, Daejeon 34134, Republic of Korea*

*[2]Research Center for Electronic and Optical Materials, National Institute for Materials Science, 1-1 Namiki, Tsukuba 305-0044, Japan.*

*[3]Research Center for Materials Nanoarchitectonics, National Institute for Materials Science, 1-1 Namiki, Tsukuba 305-0044, Japan.*

*[4]Institute for Materials Research, Tohoku University, 2-1-1 Katahira, Aoba-ku, Sendai, 980-8577, Japan*

[†] These authors equally contributed to this work.

* To whom correspondence should be addressed.

E-mail: bae.soungmin.d6@tohoku.ac.jp (S.B); kyee@cnu.ac.kr (K.J.Y.)

**ABSTRACT**

Photodoping plays an important role in determining the optoelectronic response of two-dimensional semiconductor devices; however, the origin of the responsible trap states remains unclear. In this work, we investigate UV-induced photodoping in multilayer $WSe_2$ field-effect transistors (FETs) based on $WSe_2$/hBN heterostructures on $SiO_2$/p-Si substrates. Wavelength-dependent measurements reveal pronounced n-type photodoping under 405 nm illumination, whereas the effect is orders of magnitude weaker under 640 nm excitation and for p-type photodoping. Furthermore, when the $SiO_2$ layer is removed, both n-type and p-type photodoping are strongly suppressed, demonstrating that the oxide layer is essential for persistent photodoping. Analysis of defect-state distributions in amorphous $SiO_2$, together with first-principles calculations for hBN defects, shows that the experimental observations cannot be explained by defects in hBN. Instead, the results indicate that defect states in the $SiO_2$ substrate act as charge reservoirs that facilitate charge transfer and long-term carrier trapping. These findings highlight the dominant role of substrate-related trap states in UV-induced photodoping and photogating behavior in $WSe_2$ FET devices.

## 1. Introduction

Two-dimensional (2D) layered materials, whose intrinsic properties are preserved down to the atomic thickness, have attracted increasing attention as a route to extend transistor scalability beyond bulk materials, enabling nanoelectronics and optoelectronics at the nanoscale.[1–3] Among these materials, transition metal dichalcogenides (TMDs), such as $MoS_2$ and $WSe_2$, provide a versatile platform that exhibits many exotic phenomena, including thickness dependent bandgaps, strong light–matter interactions even at the monolayer limit, and high on–off ratio transistor operation. Moreover, novel 2D heterostructure architectures with enhanced performance can be realized by combining insulating hexagonal boron nitride (hBN) or metallic graphene with semiconducting TMDs. As a result, TMDs are highly promising for optoelectronic device applications such as photodetectors, light-emitting diodes, and phototransistors.[4–6]

From simple p–n diodes to integrated circuits, the manipulation of electrical properties through doping is crucial for enabling the functionalities of 2D materials. Carrier doping in TMDs is achievable through the charge transfer of surface adatoms [7,8], dopant atom intercalation or substitution [9–11], as well as through intrinsic defects such as vacancies and structural deformations.[12–14] Charged boron vacancies with optically addressable spin properties, which can be created by ion implantation in insulating hBN, have also attracted considerable interest for integrated quantum information applications.[15–18] More specifically, substitutional p-type doping of $WSe_2$ has been demonstrated by replacing transition metal atoms with Nb to form new chemical bonds,[19] while n-type doping of $WSe_2$ has been achieved via surface charge transfer using potassium dopants, yielding a sheet carrier density of $2.5 \times 10^{12}$ $cm^{-2}$.[20] However, substitutional doping approaches are generally irreversible, and molecular doping often suffers from limited environmental stability.[21]

Moreover, ion implantation doping, although well established for silicon, is poorly compatible with atomically thin 2D lattices due to the possibility of crystalline damage.

In heterostructures with type-II band alignment, electrons and holes tend to migrate to their respective lower-energy states, resulting in delayed recombination of spatially separated carriers and modulated carrier doping.[22–25] Such spatial charge separation can also arise from the built-in electric field at the heterointerface or from an externally applied bias field. Furthermore, delayed recombination and doping effects can be further enhanced when the separated charges are trapped at energetically or spatially isolated defect states. This charge separation and trapping of photogenerated electron–hole pairs can give rise to novel optoelectronic functionalities, such as photodoping, persistent photoconductivity, enhanced photocatalysis, and gate-material induced photoresponse.[26–31]

Recently, UV photodoping in van der Waals heterostructures composed of transition metal dichalcogenides (TMDs) and hexagonal boron nitride (hBN) has attracted considerable attention as a promising approach for achieving rewritable and on-demand doping. In this method, charge separation is induced by the gate electric field in field-effect transistors (FETs). Owing to its capability to provide tunable doping levels, polarity switching between n- and p-type conduction, and precise spatial control through laser spot positioning, UV photodoping is expected to play an important role in advanced two-dimensional material applications, including optical memory devices, optoelectronic synapses, and highly sensitive photodetectors. Despite these advantages, the fundamental mechanism underlying photodoping remains incompletely understood. Various mechanisms have been suggested and are still under active investigation (Table S1, Supporting Information). One proposed mechanism involves optical transitions of donor-like defect states in hBN, which can lead to charge trapping and persistent doping in graphene/hBN or TMD/hBN heterostructures.[32–34] Alternatively, a semiconductor-channel-mediated photodoping mechanism has been suggested, in which photoexcited carriers generated in the semiconductor channel (e.g., monolayer $MoS_2$) are transferred to localized defects in hBN.[35]

In this work, we report systematic photodoping of both electron and hole carriers in a $WSe_2$/hBN heterostructure FET, controlled by the photon energy and gate voltage. The degree of photodoping can be precisely tuned by adjusting the gate bias from −60 to 60 V, achieving carrier concentrations of up to $3.3 \times 10^{12}$ $cm^{-2}$. Measurements performed while varying the exposure time and photon energy reveal that n-type photodoping is three orders of magnitude more efficient than p-type photodoping, with 405 nm excitation producing significantly stronger doping compared to 640 nm excitation. Notably, the photodoping rate does not scale with hBN thickness, a finding inconsistent with the mechanism based on optical excitation of defect states within hBN. The experimental observations are further supported by first-principles calculations, which likewise rule out lattice defects in hBN as the primary photodoping source. Instead, we propose that photodoping originates from optically activated charge transfer from defect states in the $SiO_2$ substrate through the hBN layer into the $WSe_2$ channel.

## 2. Results and Discussion

To investigate photoinduced carrier doping in the $WSe_2$/hBN heterostructure, a field-effect transistor (FET) device was fabricated using a 2D material flake-transfer process combined with conventional lithographic patterning of metal electrodes. Figure 1A shows a schematic illustration of the device structure under laser illumination, while Fig. 1B presents an optical microscope image of the fabricated device. The thicknesses of the bottom hBN and $WSe_2$ layers were determined to be 47 nm and 12 nm, respectively, from atomic force microscopy (AFM) height profiles (Fig. S1, Supporting Information). The Raman spectrum shown in Fig. S2 (Supporting Information) confirms the high crystalline quality of the $WSe_2$ channel after device fabrication.

The electrical characteristics of the FET device were measured using a dual-channel source meter (Keithley 2502) and a microprobe system under vacuum. A gate bias was applied to the p-Si substrate after partially removing the bare $SiO_2$ layer to allow electrical contact with a probe tip. Figure 1C plots the drain–source current ($I_{DS}$) as a function of gate voltage ($V_G$) at a fixed drain–source voltage ($V_{DS}$) of 0.2 V. The transfer curve indicates electron-dominated conduction with a threshold gate voltage ($V_{Th}$) of 28 V, while the logarithmic plot in the inset reveals significantly lower hole currents. The negligible hysteresis between the forward and backward gate sweeps suggests a limited contribution from oxide or surface defects after hBN encapsulation and ensures good experimental reproducibility. The electron mobility ($\mu$) of 49 $cm^2/V{\cdot}s$ is extracted from the transconductance ($g_m$) of 0.13 μA/V using the relation $g_m = CwV_{DS}\mu/L$, where $C$ = 9.75 $nF/cm^2$ is the series capacitance of the $SiO_2$ and bottom hBN layers, and w and L denote the channel width and channel length, respectively.[36,37] The current–voltage characteristics of the $WSe_2$ channel at the drain and source contacts under different $V_G$ are presented in Fig. 1D, showing quasi-ohmic behavior of the Cr/Au contacts on multilayer $WSe_2$.

To investigate the photodoping of the multilayer $WSe_2$/hBN heterostructure, laser diodes with wavelengths of 405 nm or 640 nm were focused onto the $WSe_2$ channel through a 50× objective lens. Hereafter, $V_{Ex}$ denotes the gate voltage applied during laser illumination to induce photodoping, whereas $V_G$ represents the gate bias used to measure the electrical properties after the illumination is turned off. Figure 2A shows the transfer curves measured after 405 nm illumination under various $V_{Ex}$ conditions, where the laser power and exposure duration were fixed at 144 μW and 50 s, respectively. The results clearly demonstrate that the transfer curves shift systematically along the horizontal axis depending on the gate bias applied during laser exposure. Because the channel conductivity is proportional to the total carrier density, which consists of carriers induced by electrical gating and the intrinsic carrier density in the absence of gate bias, the horizontal shift of the transfer curves indicates a modulation of the carrier density after laser exposure.

The threshold voltage in Fig. 2B, defined as the $V_G$ at which the D-S current reaches 0.3 μA, increases linearly with $V_{Ex}$ with a slope close to unity. This linear dependence indicates that the carrier density generated during UV photodoping is approximately equal to that induced by electrical gating prior to photodoping. This suggests that additional charges are created during photodoping to compensate for the gate-induced carriers initially present in the $WSe_2$ channel. As a result, substantial carrier density modulation is achieved, reaching a sheet carrier density (σ) of $3.7 \times 10^{12}$ cm$^{-2}$ for a $V_{Th}$ shift of 60 V, as estimated using the relation $\sigma = CV$. Furthermore, the semi-logarithmic plot of the transfer curve (Fig. S2, Supporting Information) shows that UV photodoping also modifies the hole current at $V_G < V_{Th}$, in addition to influencing the electron conduction.

In order to evaluate the photodoping efficiency, we investigated the exposure duration and laser wavelength dependence, of which the results are presented in Fig. 3. Prior to each measurement, the multilayer $WSe_2$ channel was initialized to a zero-gate-doping condition by illuminating it with a 405 nm laser at $V_{Ex} = 0$ V for a sufficiently long duration. Photodoping was then performed using either a 405 nm (3.06 eV) or 640 nm (1.94 eV) laser with an optical power of 17 μW for various exposure times. The illumination was carried out under $V_{Ex} = -50$ V or of $V_{Ex} = +50$ V to induce electron doping or hole doping, respectively. After each light exposure, the transfer curve was measured to quantify the degree of photodoping.

Figure 3A presents the transfer curves obtained after 405 nm laser illumination at $V_{Ex} = -50$ V for n-type photodoping, with the exposure time varied up to 50 s. The negative horizontal shift of the transfer curves, while maintaining their overall shape, indicates gradual electron doping in the $WSe_2$ channel with increasing exposure time. The shift eventually saturates at longer exposure times when the photoinduced charge distribution replaces the gate-induced carriers in the $WSe_2$ channel. Figure 3B shows the transfer curves obtained after 405 nm illumination at $V_{Ex}$ = +50 V for p-type photodoping. In contrast to the rapid n-type doping, the positive horizontal shift observed in the p-type photodoping case is significantly slower and smaller, even after an extended exposure time of 200 s.

In addition to the exposure duration, the photon energy is another important parameter governing the photodoping process. Figures 3C and 3D plot the horizontal shift of the transfer curves as a function of exposure time under 640 nm illumination for n-type and p-type photodoping, respectively. As shown in the results, the n-type photodoping rate induced by the 640 nm laser is three orders of magnitude slower than that observed with the 405 nm laser, requiring approximately 200 s to reach the midpoint of full n-type photodoping. The p-type photodoping process under 640 nm illumination is even slower, with only about 5% of the saturation level achieved after 200 s of exposure. These results clearly demonstrate that the photodoping efficiency strongly depends on both the gate polarity and the photon energy.

Figure 3E summarizes the exposure-time-dependent $V_{Th}$ shifts for n-type (dark and light blue curves) and p-type (dark red and magenta curves) photodoping. The $V_{Th}$ shift from 28 V to −21 V for n-type doping under 405 nm excitation corresponds to the applied gate bias ($|V_{Ex}|$ = 50 V), indicating saturated photodoping. Under 405 nm illumination, the n-type photodoping reaches its midpoint ($V_{Th}$ = 4 V) within 120 ms. In contrast, the positive shift of the transfer curve associated with p-type photodoping occurs much more slowly, requiring approximately 100 s to produce a comparable $V_{Th}$ change. This result indicates that p-type photodoping under 405 nm illumination is roughly three orders of magnitude slower than the n-type process. A similar polarity dependence is also observed under 640 nm illumination, where p-type photodoping remains negligible even after 200 s of exposure.

For the case of p-type photodoping, in which the laser illumination is applied while the FET is in the on-state under a positive gate bias, the decay of the electron density in the $WSe_2$ channel during the photodoping process can be traced by monitoring the channel conductivity. Figure 4A presents the temporal evolution of the drain current while the $WSe_2$ is exposed to a 405 nm laser with various optical powers under a gate voltage of +50 V. Because the conductivity $\sigma_n$ is linearly proportional to the electron density $n$ according to $\sigma_n = ne\mu$, the decay of the drain current corresponds to a decrease in the electron density in the $WSe_2$ channel, which effectively results in hole doping when the gate bias is removed. As indicated by the solid lines, the current decay can be well fitted by a stretched-exponential function, $I(t) = I_0 \exp[-(t/\tau)^{\beta}]$, where $\tau$ is the decay time constant and $\beta$ is the stretching parameter.[35] The extracted fitting parameters are plotted in Fig. 4B, where $\beta$ ranges from 0.4 to 0.6, and the decay rate $(1/\tau)$ is linearly proportional to the photon density. This proportionality between the photodoping rate and photon flux indicates systematic controllability of the doping efficiency. We note that similar behavior is expected for n-type photodoping as well; however, the negligible p-channel conductance prevents direct observation of the process through channel conductance measurements.

The experimental results confirm that photodoping in $WSe_2$/hBN heterostructures becomes significantly stronger under shorter-wavelength illumination, particularly for 405 nm light compared with 640 nm. Most previous studies on photoinduced doping in TMD/hBN heterostructures have attributed this phenomenon to optical transitions from localized defect states in the hBN bulk to delocalized bands,[28,32] while another study has suggested that optical absorption in the TMD channel itself plays the dominant role.[32,35] To clarify the mechanism responsible for the photodoping observed in our device, we critically examine the validity of the defect-mediated mechanism associated with lattice defects in hBN.

In the conventional defect-mediated model, the photodoping efficiency is expected to scale with the number of relevant defects in the hBN layer or with the hBN thickness. To verify this assumption, we performed photodoping experiments on devices with different hBN thicknesses; however, no significant variation was observed (Fig. S4, Supporting Information). This result contradicts the hBN defect-mediated photodoping mechanism and suggests that alternative mechanisms must be considered.

To examine the possible role of defects in the $SiO_2$ layer during the photodoping process, we fabricated control devices, one with $SiO_2$ layer and another with removing the $SiO_2$ layer via wet etching and compared the photodoping behavior under identical illumination conditions. Figure 5 shows the time evolution of the normalized threshold-voltage shift under both n-type and p-type photodoping conditions, where triangular symbols represent n-type doping and circular symbols represent p-type doping. The transfer-curve shifts as a function of 405 nm exposure time for both n-type and p-type photodoping are presented in Fig. S5 (Supporting Information). For devices with an intact $SiO_2$ substrate, a pronounced and time-dependent photodoping behavior is observed for both doping polarities.

In contrast, when the $SiO_2$ layer is removed, both n-type and p-type photodoping are strongly suppressed. The corresponding data (triangles and circles without $SiO_2$) exhibit negligible variation over the entire exposure time range, indicating that neither electron nor hole doping can be effectively accumulated in the absence of the $SiO_2$ layer. This clear distinction demonstrates that the presence of $SiO_2$ is essential for enabling persistent photodoping. In particular, the pronounced time-dependent evolution observed only in $SiO_2$-supported devices suggests that oxide-related defect states act as a charge reservoir, facilitating the accumulation and long-term retention of photoinduced carriers. These results provide direct evidence that photodoping is critically mediated by charge transfer and trapping at defect states in $SiO_2$, rather than originating from point defects in the $WSe_2$ or hBN layers.

To better understand the role of $SiO_2$ in photodoping, we consider the nature of defect states in amorphous $SiO_2$, which is known to host both shallow and deep trap states for electrons and holes that can serve as charge reservoirs required for photodoping. The estimated positions of these defect states within the energy gap are schematically illustrated in Fig. 6A. Electron trapping in amorphous $SiO_2$ can arise from intrinsic structural distortions of the amorphous network, leading to deep trap states located approximately 2.8−3.0 eV below the conduction band minimum. Shallow electron traps, associated with oxygen-vacancy−related configurations, form a broad distribution at 0.8−1.4 eV below the conduction band.[36,37] In comparison, hole-related states exhibit a more asymmetric distribution, consisting of shallow states near the valence band maximum, often associated with $E'_{\delta}$ dimer configurations, and deep trap states located approximately 4.3–4.6 eV above the valence band maximum, which are attributed to $E_{\gamma}'$ puckered configurations.[39,40]

Considering the broad energy distribution of these trap states, the activation of deep electron traps by 405 nm photons is expected to be efficient, whereas the activation of deep hole traps under 405 nm excitation and that of deep electron traps under 640 nm excitation are expected to be significantly less efficient. The trap-state activation pathways enabled by 405 nm or 640 nm photons, which lead to charge carrier transport into the $WSe_2$ channel, are schematically illustrated in Fig. 6B. Although shallow trap states are expected to be optically accessible under both 405 nm (3.06 eV) and 640 nm (1.94 eV) excitation, the experimentally observed photodoping under 640 nm illumination is orders of magnitude weaker. Together with the observed polarity dependence, this result suggests that shallow traps contribute less effectively to the photodoping behavior than deep trap states, indicating that higher-energy excitation is required for efficient photodoping.[41] It is noteworthy that shallow trap states, such as $E'_{\delta}$ configurations, are generally associated with short trapping lifetimes and rapid thermal release of carriers, making them unsuitable for sustaining the long-lived charge accumulation required for persistent photodoping. In contrast, deep trap states, including $E_\gamma'$ centers, can provide longer retention times and thus enable the accumulation of stable photoinduced charge.[37,39]

We additionally performed Kelvin probe force microscopy (KPFM) measurements on a test FET device without top hBN capping in order to confirm the transferred charge carriers after the photodoping process. As shown in Fig. S6 (Supporting Information), the work function of $WSe_2$ exhibits a positive (negative) shift after n-type (p-type) photodoping, respectively. These results provide direct evidence of charge carrier generation in the $WSe_2$ channel induced by the photodoping process. The persistence of the photodoping effect is also presented in Fig. S7 (Supporting Information), where the n-type (p-type) carriers generated by photodoping decrease by only 6.4% (4.6%) after 10 hours.

The role of bulk defects in hBN as the primary source of photodoping is further weakened by first-principles calculations. Details of the calculations are provided in the Supporting Information. As shown in Fig. 6C, several acceptor-like defect states in hBN are energetically accessible even under 640 nm excitation; however, the experimentally observed p-type photodoping under this condition is negligible. This inconsistency further disfavors hBN defects as the dominant origin of photodoping and instead supports defect states in $SiO_2$ as the primary activation source of UV-induced photodoping in $WSe_2$ FETs.

Defects in $WSe_2$ may also contribute to the photodoping effect if photogenerated carriers are trapped at these defect sites for extended periods. However, the negligible photodoping observed in $WSe_2$/hBN heterostructures without a $SiO_2$ layer, in contrast to the pronounced photodoping in devices incorporating $SiO_2$, indicates that defects in $SiO_2$ play a dominant role, with only a minor contribution, if any, from $WSe_2$ defects. We attribute this behavior to either a relatively low intrinsic defect density in $WSe_2$ or to fast detrapping dynamics of carriers at $WSe_2$ defect sites.

## 3. Conclusion

In summary, we have demonstrated controllable photodoping in a multilayer $WSe_2$/hBN heterostructure field-effect transistor through the combined effects of photon energy and gate electric field. Systematic measurements reveal that the photodoping process strongly depends on both the illumination wavelength and the polarity of the applied gate bias. In particular, n-type photodoping under 405 nm excitation occurs three orders of magnitude faster than p-type photodoping, while photodoping under 640 nm illumination is significantly suppressed for both polarities. The resulting threshold-voltage shift enables modulation of the carrier density up to $\sim 3.7 \times 10^{12}$ cm$^{-2}$, demonstrating substantial and controllable doping of the $WSe_2$ channel.

Through a series of control experiments and mechanistic analyses, we show that the photodoping behavior cannot be explained by defect states in the hBN layer. The photodoping efficiency exhibits no measurable dependence on hBN thickness, and first-principles calculations indicate that hBN defect states should remain optically accessible even under conditions where photodoping is experimentally negligible. Instead, the strong suppression of photodoping after removal of the $SiO_2$ layer provides direct evidence that defect states in the oxide substrate play a crucial role in the photodoping process. Based on these observations, we propose that photodoping originates from optically activated charge transfer from trap states in $SiO_2$ through the hBN layer into the $WSe_2$ channel, where deep trap states in the oxide act as long-lived charge reservoirs enabling persistent carrier accumulation.

These findings clarify the microscopic origin of UV-induced photodoping in $WSe_2$/hBN heterostructure devices and highlight the critical role of substrate defect states in governing photoinduced carrier dynamics in van der Waals heterostructures. The ability to achieve reversible and spatially controllable carrier doping through optical excitation provides a promising strategy for implementing rewritable doping patterns in two-dimensional materials, which may enable future applications in optoelectronic memory devices, programmable phototransistors, and neuromorphic optoelectronic systems.

## METHODS

**Fabrication of $WSe_2$/hBN field-effect transistor.** Initially, hBN and $WSe_2$ flakes were mechanically exfoliated onto sacrificial Si substrates, after which the bottom hBN and $WSe_2$ layers were sequentially transferred onto a p-type Si substrate with a 300 nm $SiO_2$ layer. To minimize polymer residues, the polypropylene carbonate (PPC) dry transfer method was employed during the pickup and release process.[41] Drain and source metal electrodes were then patterned on the $WSe_2$ channel using e-beam lithography on a spin-coated PMMA resist layer, followed by Cr/Au (4/50 nm) deposition. Finally, the device was encapsulated with a top hBN layer to suppress ambient molecular adsorption and improve long-term stability.

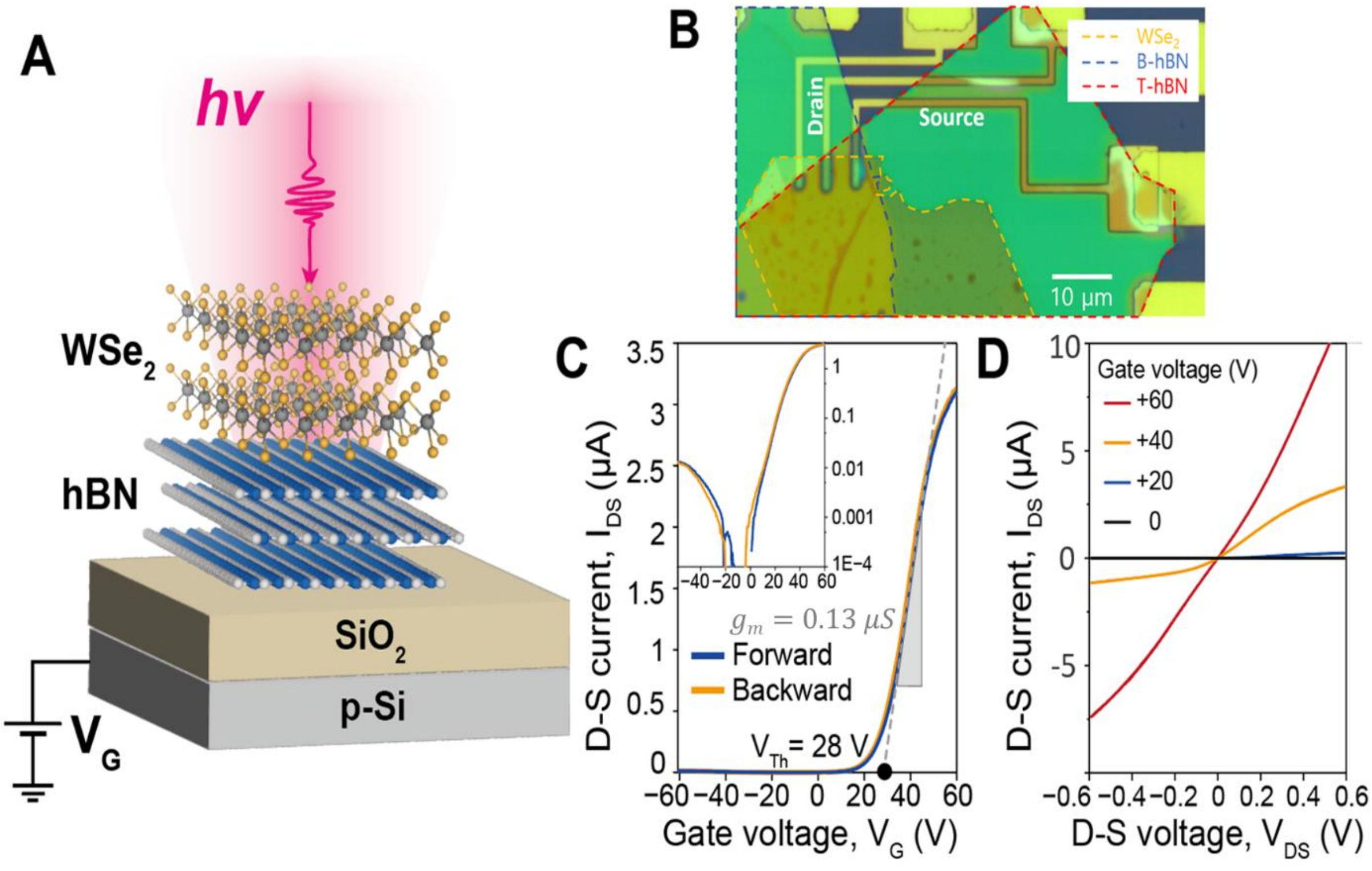

Figure 1. A) Schematic illustration of the $WSe_2$/hBN heterostructure field-effect transistor (FET) used for photodoping under gate bias and illumination. B) Optical microscope image of the fabricated device, where B-hBN is for bottom hBN and T-hBN for top hBN. C) Drain–source current ($I_{DS}$) as a function of gate voltage ($V_G$), with the corresponding semi-log plot shown in the inset. D) Drain–source current–voltage ($I_{DS}$–$V_{DS}$) characteristics of the device measured under different gate voltages ($V_G$).

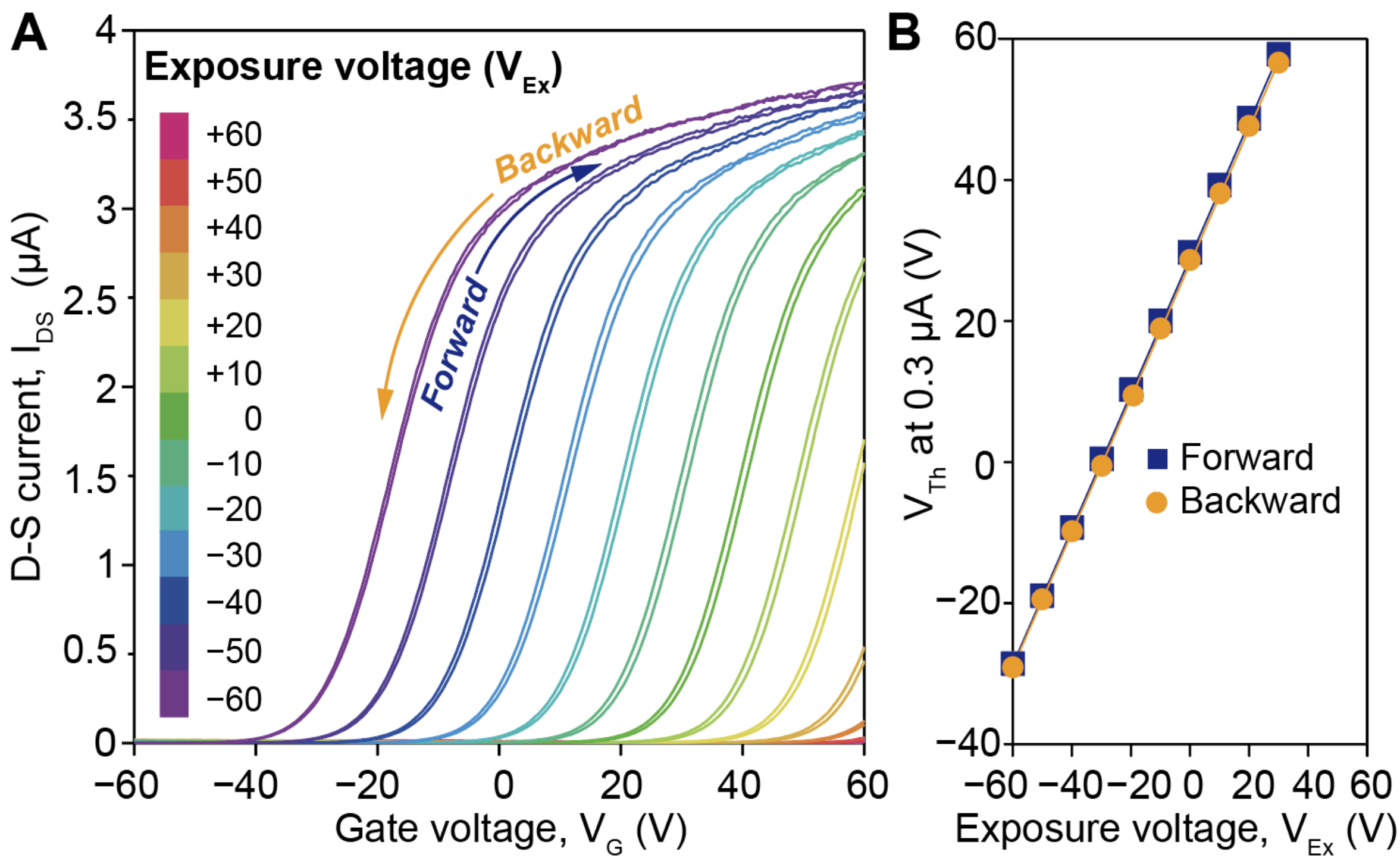

Figure 2. A) Photodoping induced shift of the transfer curve obtained after 405 nm laser exposure at different gate voltages for a 50 s duration and optical power of 144 μW. B) Threshold voltage defined as the point of $I_{DS}$ = 0.3 A as a function of the exposure voltage ($V_{Ex}$) for the forward and backward sweep.

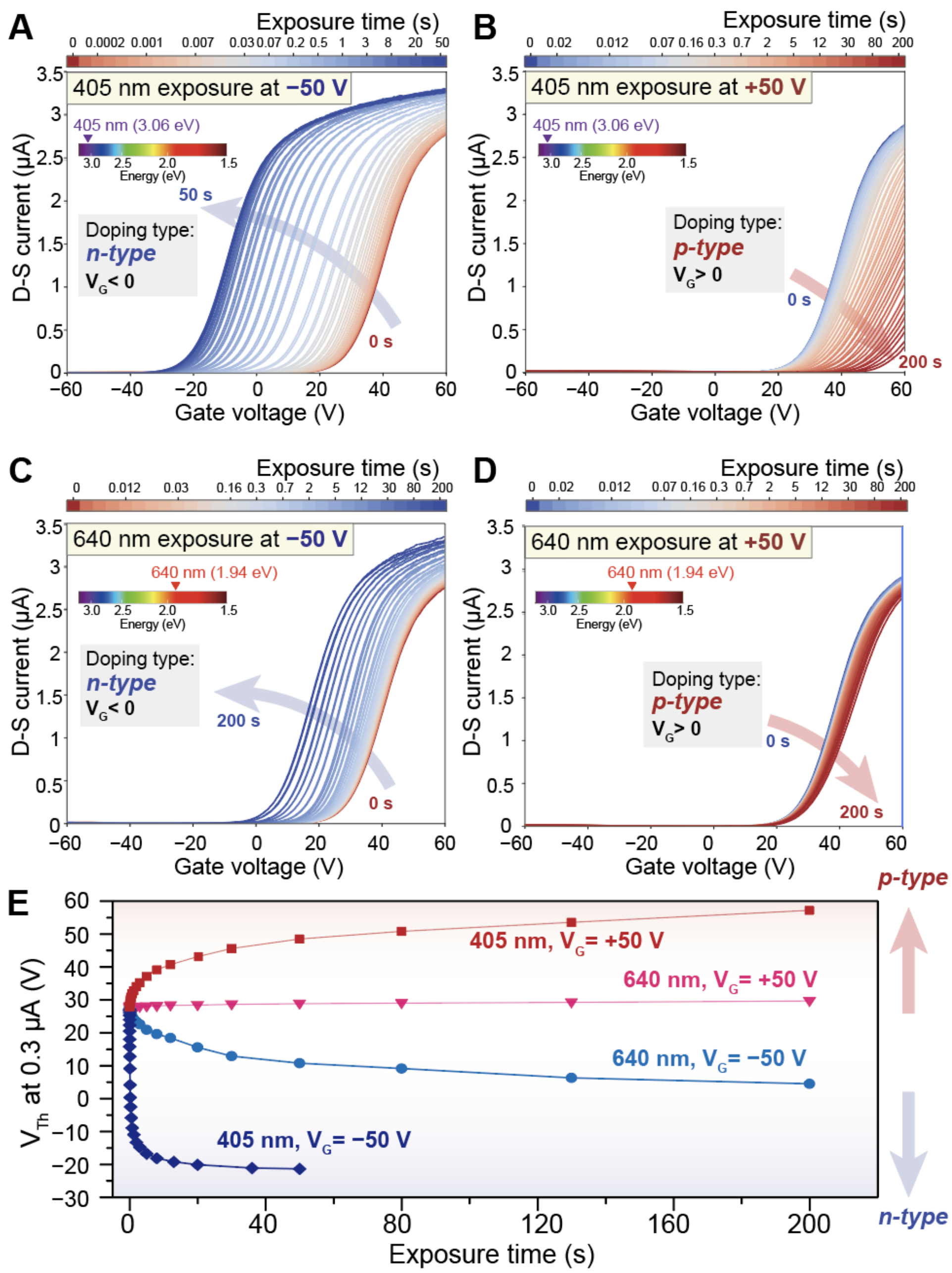

Figure 3. Exposure time dependence of the transfer curve obtained after laser exposure of different wavelengths at negative and positive gate biases: A) 405 nm exposure at −50 V, B) 405 nm exposure at 50 V, C) 640 nm exposure at −50 V, and D) 640 nm exposure at 50 V. A fixed laser power of 17 µW applies to all cases. E) Threshold voltage of the n-type transport, defined as the point where the drain current crosses 0.3 µA, as a function of exposure time for 405 nm or 640 nm wavelength and an exposure bias of +50 V for p-type or −50 V for n-type doping.

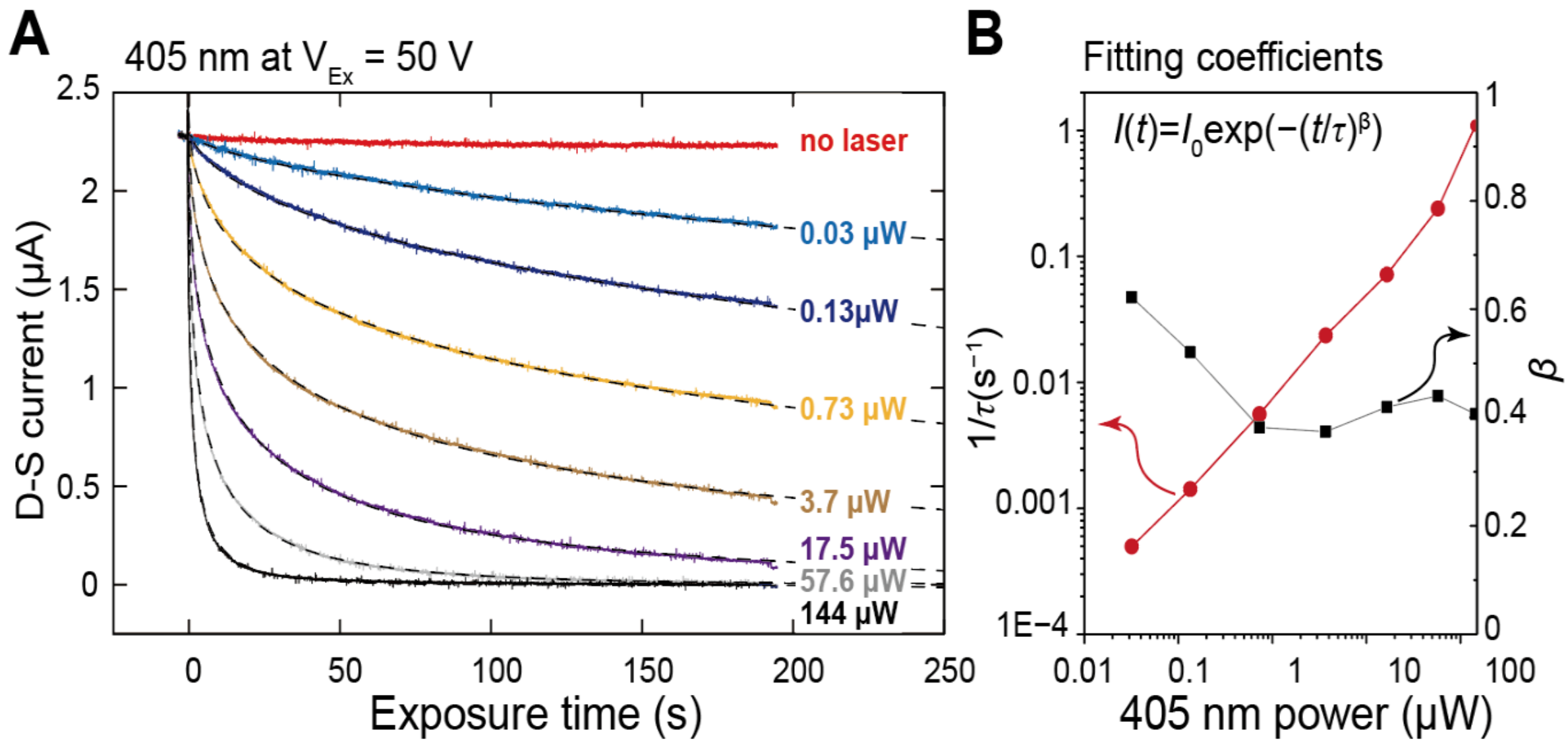

Figure 4. A) Temporal evolution of the drain current as a function of 405 nm laser exposure time of various powers at a gate bias of 50 V for p-type photodoping. The red line is the case without laser exposure. The dashed lines are fitting curves of temporal evolution with the stretched-exponential function of $f(t) = I_0\exp[-(t/\tau)^\beta]$. B) Fitting parameters $\tau$ and $\beta$ as a function of 405 nm laser power.

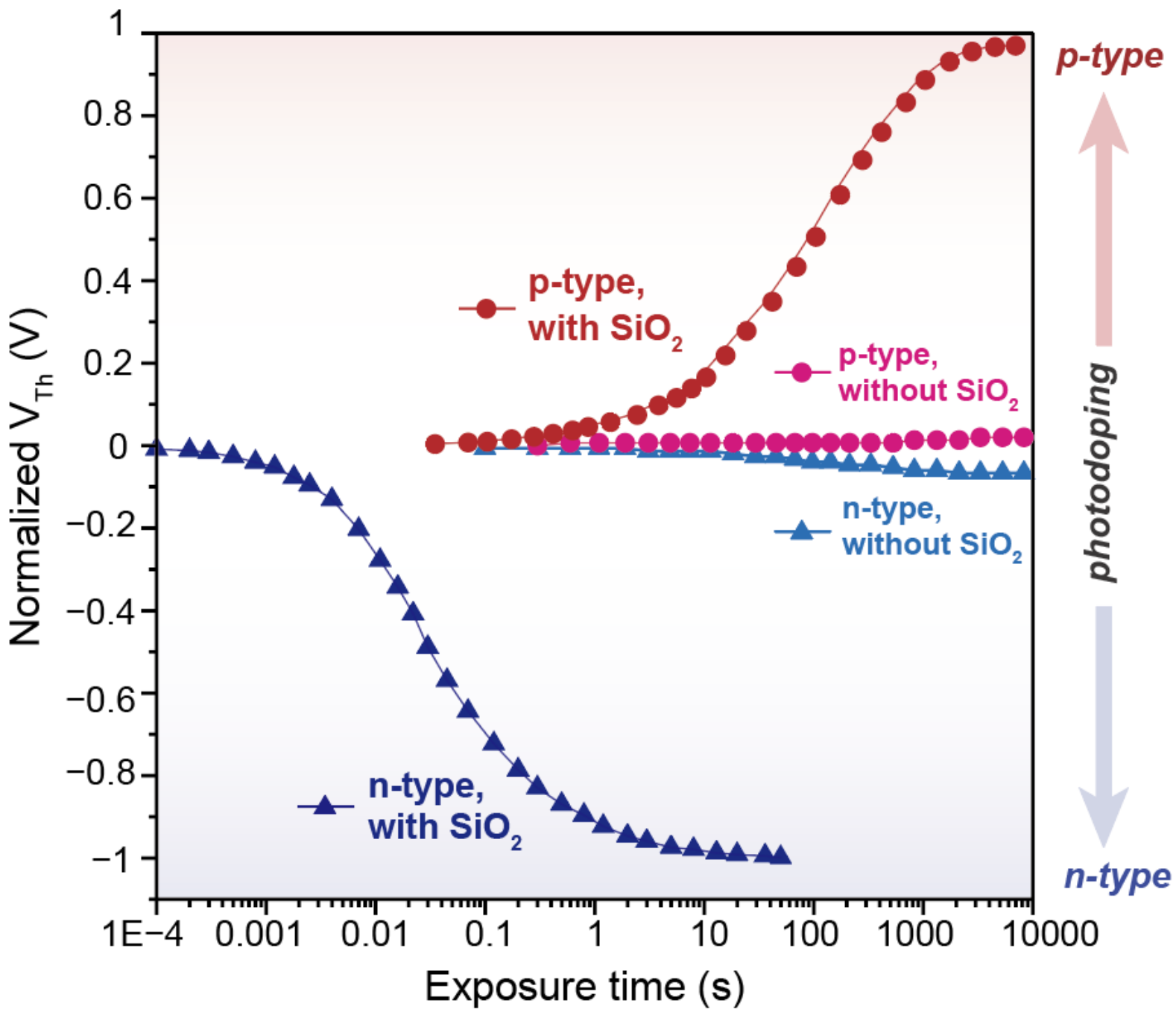

Figure 5. Role of the $SiO_2$ substrate in photodoping. Time-dependent evolution of the normalized threshold voltage shift ($\Delta V_{Th}$) under illumination for devices with and without the $SiO_2$ substrate. The $SiO_2$ layer was removed by wet etching for the control device. Triangular symbols represent n-type photodoping, while circular symbols represent p-type photodoping.

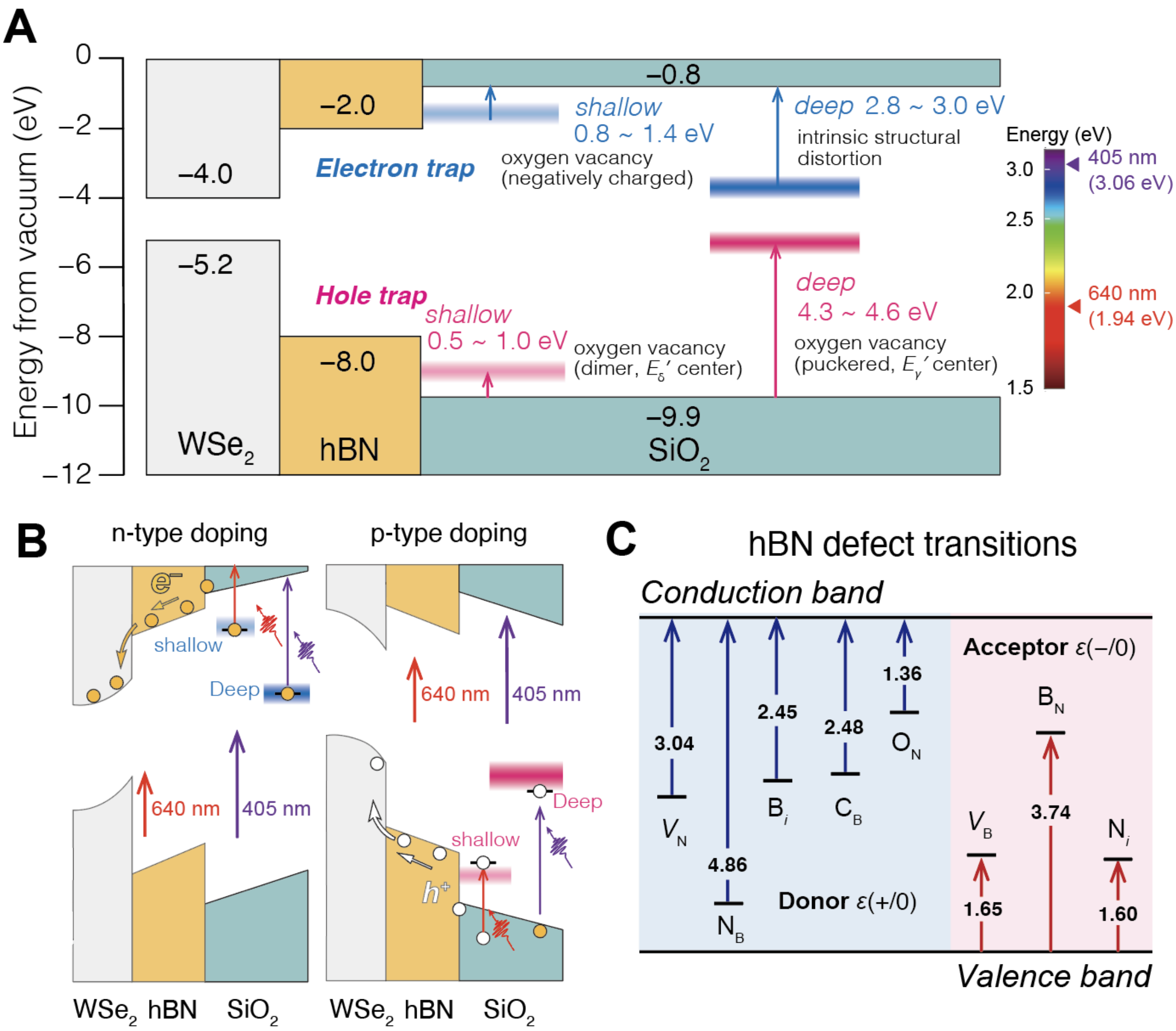

Figure 6. Origin of asymmetric photodoping in $WSe_2$/hBN/$SiO_2$ heterostructures. (A) Vacuum-level-aligned band diagram showing electron and hole trap states in $SiO_2$ that can act as charge reservoirs for photodoping, including shallow and deep levels. (B) Schematic of the photodoping mechanism. Carriers are transferred to and trapped at the hBN/$SiO_2$ interface, where deep electron traps enable efficient and persistent n-type photodoping, while shallow, short-lived hole traps limit p-type photodoping. (C) DFT-calculated excitation energies of hBN defect states, suggesting that hBN bulk defects are unlikely to dominate the photodoping behavior.

**SUPPORTING INFORMATION AVAILABLE:** Atomic force microscope height profiles of the $WSe_2$ and hBN flakes, semi-log plot of the transfer curve after the photodoping process, photodoping efficiencies on devices with different hBN thicknesses. This material is available free of charge via http://pubs.acs.org.

## ACKNOWLEDGEMENTS

This research was supported by the National Research Foundation of Korea (NRF-2020R1A6A1A03047771, NRF-2023R1A2C1004284, Creation of the quantum information science R&D ecosystem with RS-2023-00256050) and the JSPS KAKENHI (Grant No. 25K17816) from the Japan Society for the Promotion of Science. We acknowledge the Center for Computational Materials Science, Institute for Materials Research, Tohoku University (project no. 2202412-SCKXX-04071), and Supercomputer Center, Institute for Solid State Physics, University of Tokyo, for the use of the facilities. Part of the computations in this work was performed under the young scientist-user program using the miyabi-C and Wisteria/BDEC-01 Odyssey supercomputer systems at the University of Tokyo.

## REFERENCES

(1) Geim, A. K.; Novoselov, K. S. The Rise of Graphene. *Nat. Mater.* **2007**, *6* (3), 183-191.

(2) Ajayan, P.; Kim, P.; Banerjee, K. Two-dimensional van der Waals Materials. *Phys. Today* **2016**, *69* (9), 38-44.

(3) Novoselov, K. S.; Mishchenko, A.; Carvalho, A.; Castro Neto, A. H. 2D Materials and van der Waals Heterostructures. *Science* **2016**, *353* (6298), aac9439.

(4) Li, Z.; Chen, J.; Dhall, R.; Cronin, S. B. Highly Efficient, High Speed Vertical Photodiodes Based on Few-layer $MoS_2$. *2D Mater.* **2016**, *4* (1), 015004.

(5) Ross, J. S.; Klement, P.; Jones, A. M.; Ghimire, N. J.; Yan, J.; Mandrus, D. G.; Taniguchi, T.; Watanabe, K.; Kitamura, K.; Yao, W.; et al. Electrically Tunable Excitonic Light-emitting diodes Based on Monolayer $WSe_2$ p-n Junctions. *Nat. Nanotechnol.* **2014**, *9* (4), 268-272.

(6) Yin, Z.; Li, H.; Li, H.; Jiang, L.; Shi, Y.; Sun, Y.; Lu, G.; Zhang, Q.; Chen, X.; Zhang, H. Single-layer $MoS_2$ Phototransistors. *ACS Nano* **2012**, *6* (1), 74-80.

(7) Chen, W.; Qi, D.; Gao, X.; Wee, A. T. S. Surface Transfer Doping of Semiconductors. *Prog. Surf. Sci.* **2009**, *84* (9-10), 279-321.

(8) Du, H.; Zhao, G.; Liu, G.; Lv, X.; Wei, W.; Wang, X. Analysis of the Heterojunction Band Offset of h-BN/TMDCs. *Appli. Surf. Sci.* **2024**, *664*.

(9) Ye, L.; Chen, S.; Li, W.; Pi, M.; Wu, T.; Zhang, D. Tuning the Electrical Transport Properties of Multilayered Molybdenum Disulfide Nanosheets by Intercalating Phosphorus. *J. Phys. Chem. C* **2015**, *119* (17), 9560-9567.

(10) Yun, S. J.; Duong, D. L.; Ha, D. M.; Singh, K.; Phan, T. L.; Choi, W.; Kim, Y. M.; Lee, Y. H. Ferromagnetic Oder at Room Temperature in Monolayer $WSe_2$ Semiconductor via Vanadium Dopant. *Adv. Sci.* **2020**, *7* (9), 1903076.

(11) Bae, S.; Miyamoto, I.; Kiyohara, S.; Kumagai, Y. Universal Polaronic Behavior in Elemental Doping of $MoS_2$ from First-principles. *ACS Nano* **2024**, *18* (50), 33988-33997.

(12) Lin, Y.-C.; Björkman, T.; Komsa, H.-P.; Teng, P.-Y.; Yeh, C.-H.; Huang, F.-S.; Lin, K.-H.; Jadczak, J.; Huang, Y.-S.; Chiu, P.-W.; Krasheninnikov, A. V.; Suenaga, K. Three-fold rotational defects in two-dimensional transition metal dichalcogenides. *Nat. Commun.* **2015**, 6, 6736.

(13) Mitterreiter, E.; Schuler, B.; Micevic, A.; Hernangómez-Pérez, D.; Barthelmi, K.; Cochrane, K. A.; Kiemle, J.; Sigger, F.; Klein, J.; Wong, E.; Barnard, E. S.; Watanabe, K.; Taniguchi, T.; Frank, M.; Jahnke, L.; Finley, J. J.; Schwartzberg, A.; Qiu, D. Y.; Refaely-Abramson, S.; Holleitner, A. W.; Weber-Bargioni, A.; Kastl, C. The role of chalcogen vacancies for atomic defect emission in $MoS_2$. *Nat. Commun.* **2021**, 12, 3822.

(14) Żuberek, E. Z.; Olejnik, J.; Debus, J.; Ho, C.-H.; Watanabe, K.; Taniguchi, T.; Bryja, L.; Jadczak, J. Photon upconversion of defect-bound excitons in hBN-encapsulated $MoS_2$ monolayer. *J. Phys. Chem. C* **2024**, 128, 19288−19296.

(15) Guo, N. J.; Liu, W.; Li, Z. P.; Yang, Y. Z.; Yu, S.; Meng, Y.; Wang, Z. A.; Zeng, X. D.; Yan, F. F.; Li, Q.; et al. Generation of Spin Defects by Ion Implantation in Hexagonal Boron Nitride. *ACS Omega* **2022**, *7* (2), 1733-1739.

(16) Roy, T.; Tosun, M.; Kang, J. S.; Sachid, A. B.; Desai, S. B.; Hettick, M.; Hu, C. C.; Javey, A. Field-effect Transistors Built from All Two-dimensional Material Components. *ACS Nano* **2014**, *8* (6), 6259-6264.

(17) Gottscholl, A.; Kianinia, M.; Soltamov, V.; Orlinskii, S.; Mamin, G.; Bradac, C.; Kasper, C.; Krambrock, K.; Sperlich, A.; Toth, M.; et al. Initialization and Read-out of Intrinsic Spin Defects in a van der Waals Crystal at Room Temperature. *Nat. Mater.* **2020**, *19* (5), 540-545.
(18) Yin, L.; Cheng, R.; Ding, J.; Jiang, J.; Hou, Y.; Feng, X.; Wen, Y.; He, J. Two-dimensional Semiconductors and Transistors for Future Integrated Circuits. *ACS Nano* **2024**, *18* (11), 7739-7768.
(19) Pandey, S. K.; Alsalman, H.; Azadani, J. G.; Izquierdo, N.; Low, T.; Campbell, S. A. Controlled p-type Substitutional Doping in Large-area Monolayer $WSe_2$ Crystals Grown by Chemical Vapor Deposition. *Nanoscale* **2018**, *10* (45), 21374-21385.
(20) Fang, H.; Tosun, M.; Seol, G.; Chang, T. C.; Takei, K.; Guo, J.; Javey, A. Degenerate n-doping of Few-layer Transition Metal Dichalcogenides by Potassium. *Nano Lett.* **2013**, *13* (5), 1991-1995.
(21) Yoo, H.; Heo, K.; Ansari, M. H. R.; Cho, S. Recent Advances in Electrical Doping of 2D Semiconductor Materials: Methods, Analyses, and Applications. *Nanomaterials* **2021**, *11* (4), 832.
(22) Shahrokhi, M.; Raybaud, P.; Le Bahers, T. 2D $MoO_{3-x}S_x/MoS_2$ van der Waals Asembly: a Tunable Heterojunction with Attractive Properties for Photocatalysis. *ACS Appl. Mater. Interfaces* **2021**, *13* (30), 36465-36474.
(23) Ghosh, A.; Ball, B.; Pal, S.; Sarkar, P. Ultrafast Charge Transfer and Delayed Recombination in Graphitic-CN/$WTe_2$ van der Waals Heterostructure: a Time Domain Ab Initio Study. *J. Phys. Chem. Lett.* **2022**, *13* (34), 7898-7905.
(24) George, A.; Fistul, M. V.; Gruenewald, M.; Kaiser, D.; Lehnert, T.; Mupparapu, R.; Neumann, C.; Hübner, U.; Schaal, M.; Masurkar, N.; et al. Giant Persistent Photoconductivity in Monolayer $MoS_2$ Field-effect Transistors. *npj 2D Mater. Appli.* **2021**, *5* (1), 15.
(25) Arora, R.; Barr, A. R.; Larson, D. T.; Pizzochero, M.; Kaxiras, E. Engineering Interfacial Charge Transfer through Modulation Doping for 2D electronics. *Phys. Rev. Mater.* **2025**, *9* (2), L021601.
(26) Liu, H.; Yi, Z.; Chen, C. Advances in Photodetectors via Photogating Effect. *Mater. Res. Bull.* **2025**, *192*, 113628.
(27) Ju, L.; Velasco, J., Jr.; Huang, E.; Kahn, S.; Nosiglia, C.; Tsai, H. Z.; Yang, W.; Taniguchi, T.; Watanabe, K.; Zhang, Y.; et al. Photoinduced Doping in Heterostructures of Graphene and Boron Nitride. *Nat. Nanotechnol.* **2014**, *9* (5), 348-352.
(28) Liu, T.; Xiang, D.; Zheng, Y.; Wang, Y.; Wang, X.; Wang, L.; He, J.; Liu, L.; Chen, W.

Nonvolatile and Programmable Photodoping in $MoTe_2$ for Photoresist-free Complementary Electronic Devices. *Adv. Mater.* **2018**, *30* (52), e1804470.
(29) Sumanth, A.; Lakshmi Ganapathi, K.; Ramachandra Rao, M. S.; Dixit, T. A Review on Realizing the Modern Optoelectronic Applications through Persistent Photoconductivity. *J. Phys. D: Appl. Phys.* **2022**, *55* (39), 393001.
(30) Wang, Y.; Wang, Q.; Zhan, X.; Wang, F.; Safdar, M.; He, J. Visible Light Driven Type II Heterostructures and Their Enhanced Photocatalysis Properties: a Review. *Nanoscale* **2013**, *5* (18), 8326-8339.
(31) Huang, T.-E.; Wang, C.-Y.; Liu, H.-H.; Liang, B.-W.; Peng, S.-C.; Huang, Y.-J.; Lan, Y.-W.; Hung, K.-M.; Lo, K. Y. Revealing the Active Role of the Gate Electrode in Weak-Light Detection. *ACS Nano* **2026**, 20 (10), 8266–8274.
(32) Xiang, D.; Liu, T.; Xu, J.; Tan, J. Y.; Hu, Z.; Lei, B.; Zheng, Y.; Wu, J.; Neto, A. H. C.; Liu, L.; et al. Two-dimensional Multibit Optoelectronic Memory with Broadband Spectrum Distinction. *Nat. Commun.* **2018**, *9* (1), 2966.
(33) Aftab, S.; Iqbal, M. Z.; Iqbal, M. W. Programmable Photo-induced Doping in 2D materials. *Adv. Mater. Interfaces* **2022**, *9* (32), 2201209.
(34) Tsai, M.-Y.; Huang, C.-T.; Lin, C.-Y.; Lee, M.-P.; Yang, F.-S.; Li, M.; Chang, Y.-M.; Watanabe, K.; Taniguchi, T.; Ho, C.-H.; et al. A Reconfigurable Transistor and Memory Based on a Two-dimensional Heterostructure and Photoinduced Trapping. *Nature Electronics* **2023**, *6* (10), 755-764.
(35) Quereda, J.; Ghiasi, T. S.; van der Wal, C. H.; van Wees, B. J. Semiconductor Cannel-mediated Photodoping in h-BN Encapsulated Monolayer $MoSe_2$ Phototransistors. *2D Mater.* **2019**, *6* (2), 025040.
(36) Hamilton, M. C.; Martin, S.; Kanicki, J. Field-effect Mobility of Organic Polymer Thin-film Transistors. *Chem. Mater.* **2004**, *16* (23), 4699-4704.
(37) El-Sayed , A.-M,; Watkins, M. B.; Afanas’ev, V. V.; Shluger A. L. Nature of Intrinsic and Extrinsic Electron Trapping in $SiO_2$, *Phys. Rev. B* **2014**, 89, 125201.
(38) Nicklaw, C. J.; Lu, Z.-Y.; Fleetwood, D. M.; Schrimp, R. D.; Pantelides, S. T. The Structure, Properties, and Dynamics of Oxygen Vacancies in Amorphous $SiO_2$, *IEEE Trans. Nucl. Sci.* **2002**, *49* (6), 2667-2673.
(39) Sushko, P. V.; Mukhopadhyay, S.; Stoneham, A. M.; Shluger, A. L. Oxygen vacancies in amorphous silica: structure and distribution of properties, *Microelectron. Eng.* **2005**, 80, 292.
(40) Yue, Y.; Song, Y.; Zuo, X. First principles study of oxygen vacancy defects in amorphous

$SiO_2$, *AIP Advances* **2017**, 7, 015309.
(41) Watson, A. J.; Lu, W.; Guimarães, M. H. D.; Stöhr, M. Transfer of Large-scale Two-dimensional Semiconductors: Challenges and Developments. *2D Mater.* **2021**, *8* (3), 032001.

Table of Contents (TOC) graphic

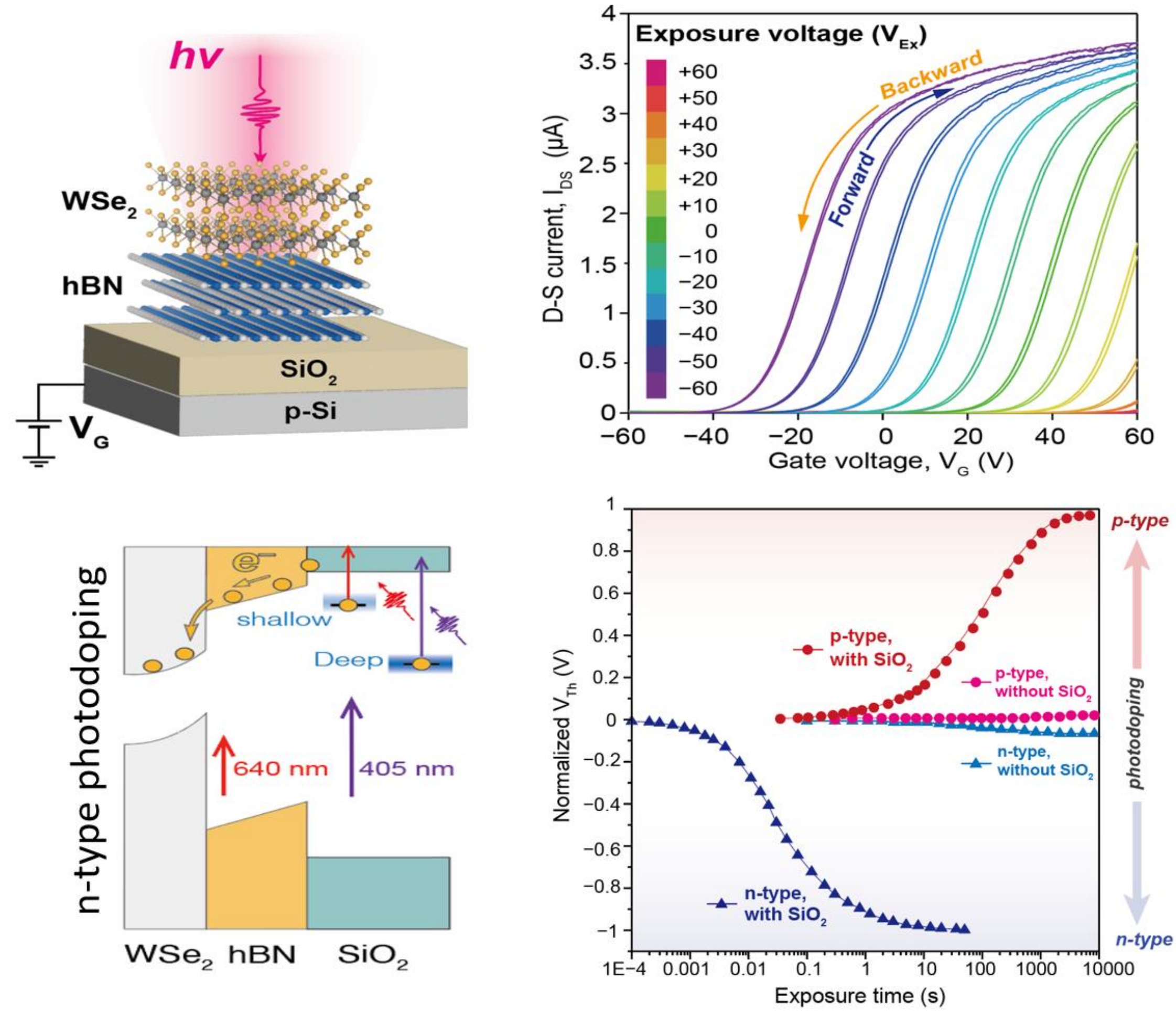

*Supporting Information for*

# Substrate-Mediated Persistent Photodoping in $WSe_2$/hBN Field-Effect Transistors Enabled by Defect States in $SiO_2$

*Sung-Ha Kim[1,†], Seong-Yeon Lee[1,†], Tae-Jeong Kim[1], Beomkyu Shin[1], Young-Jun Yu[1], Kenji Watanabe[2], Takashi Taniguchi[3], Soungmin Bae[4,*], and Ki-Ju Yee[1,*]*

*[1]Department of Physics, Chungnam National University, Daejeon 34134, Republic of Korea*

*[2]Research Center for Electronic and Optical Materials, National Institute for Materials Science, 1-1 Namiki, Tsukuba 305-0044, Japan.*

*[3]Research Center for Materials Nanoarchitectonics, National Institute for Materials Science, 1-1 Namiki, Tsukuba 305-0044, Japan.*

*[4]Institute for Materials Research, Tohoku University, 2-1-1 Katahira, Aoba-ku, Sendai, 980-8577, Japan*

[†] These authors equally contributed to this work.

* To whom correspondence should be addressed.

E-mail: bae.soungmin.d6@tohoku.ac.jp (S.B); kyee@cnu.ac.kr (K.J.Y.)

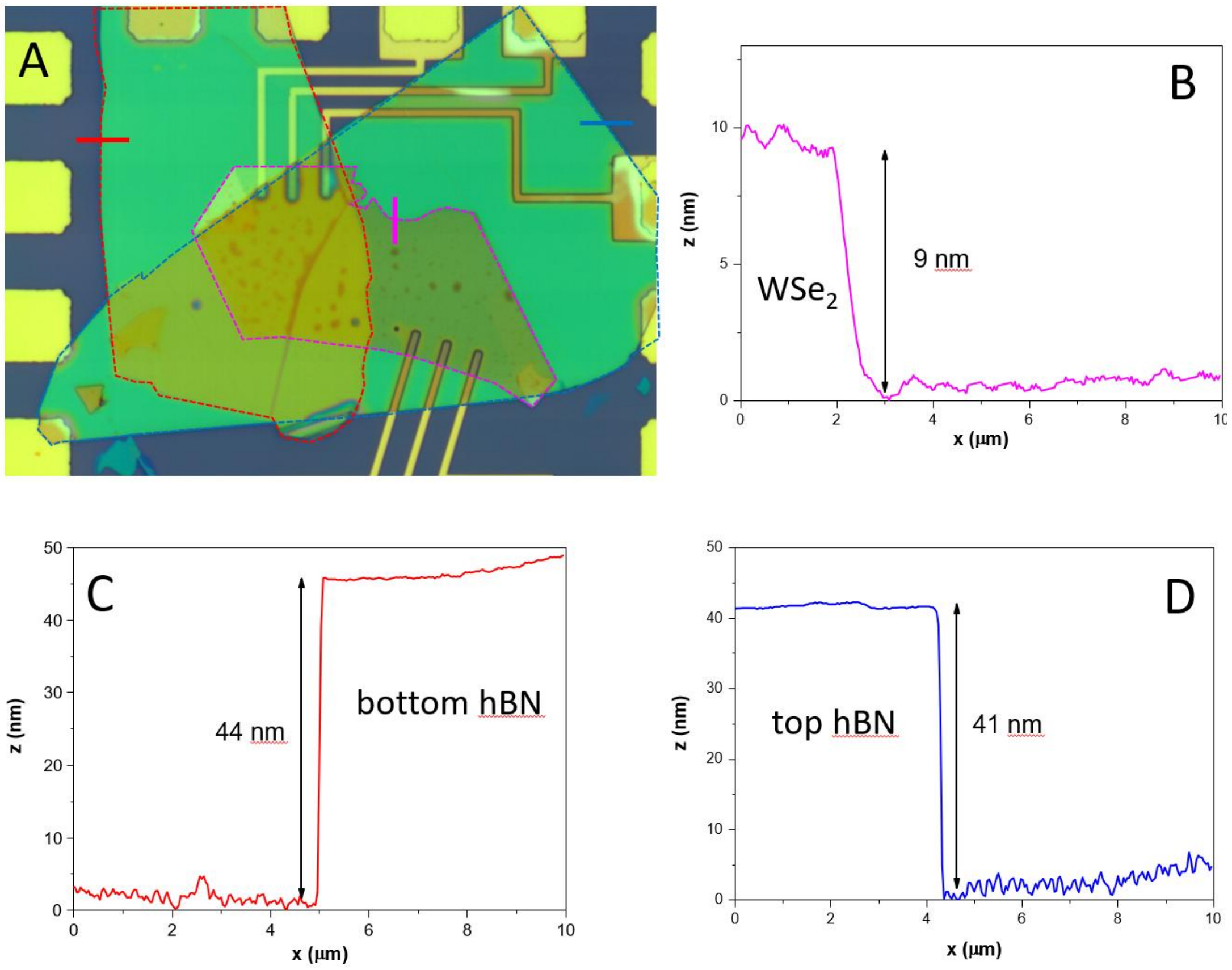

**Figure S1. Atomic force microscope height profiles of the $WSe_2$ and hBN flakes.** A) Optical microscope image of the fabricated $WSe_2$/hBN heterostructure field-effect transistor tested for the photodoping experiments. B, C, D) The atomic force microscopy line scans across the boundaries of the $WSe_2$, bottom hBN, and top hBN flakes, which confirm that the estimated thicknesses are 9 nm, 44 nm, and 41 nm, respectively.

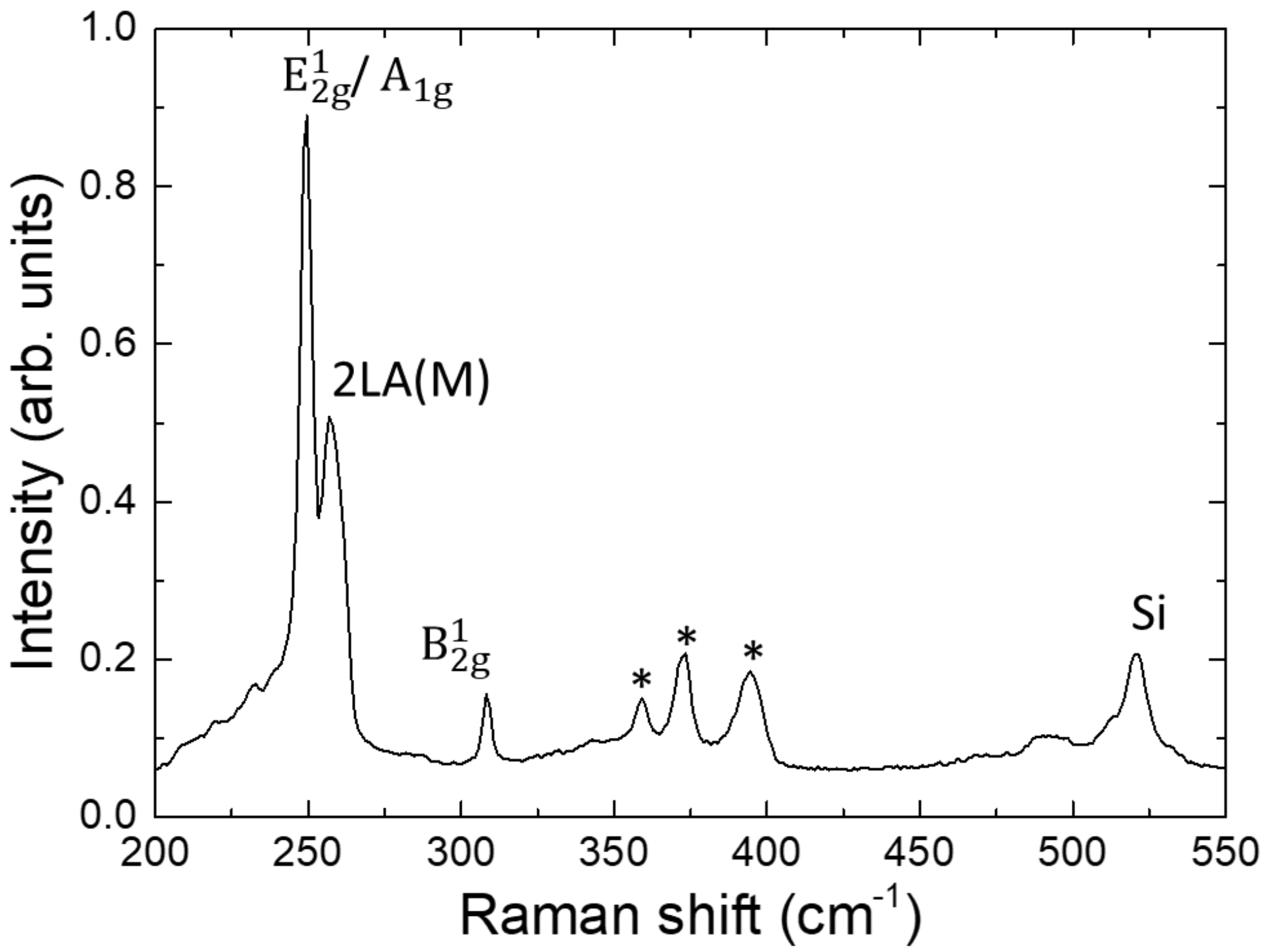

**Figure S2. Raman spectrum of multilayer $WSe_2$ channel.** Raman spectrum of the $WSe_2$ channel region obtained measured using a 532 nm excitation laser. The peak at 249 $cm^{-1}$ originates from the combined contributions of the in-plane $\boldsymbol{E_{2g}^{1}}$ mode and the out-of-plane $\boldsymbol{A_{1g}}$ mode atomic vibration, which are known to appear very close to each other in multilayer $WSe_2$. The peak at 257 $cm^{-1}$ is assigned as the 2LA(M) mode, while the peak at 308 $cm^{-1}$ corresponds to the interlayer $\boldsymbol{B_{2g}^{1}}$ mode. The three peaks between 350 $cm^{-1}$ and 400 $cm^{-1}$, marked with asterisks are attributed second-order Raman modes such as $ZO_1$(M)+TA(M) and $LO_2$(M)+LA(M). [S1, S2] The Raman peak energies agree well with previously reported values for multilayer $WSe_2$ flakes, confirming the crystalline quality of the $WSe_2$ used in this study.

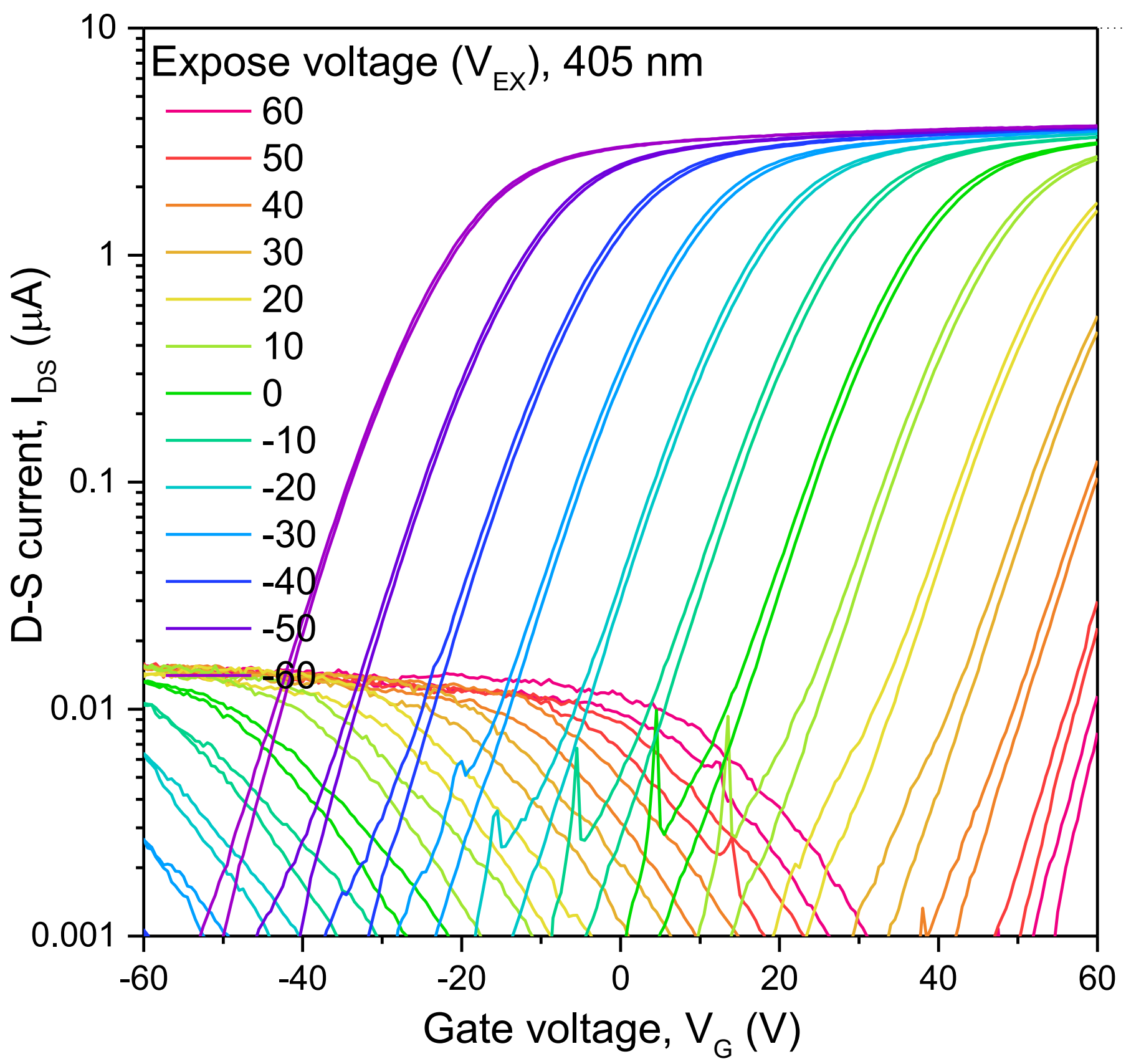

**Figure S3. Semi-log plot of the transfer curve after the photodoping process.** Photodoping induced shift of the transfer curve obtained after 405 nm laser exposure at different gate voltages for 50 s duration and optical power of 144 μW. The semi-log plot exhibits the photodoping-induced horizontal shift of the hole currents at negative gate voltages, together with that of electron currents at $V_G > V_{th}$.

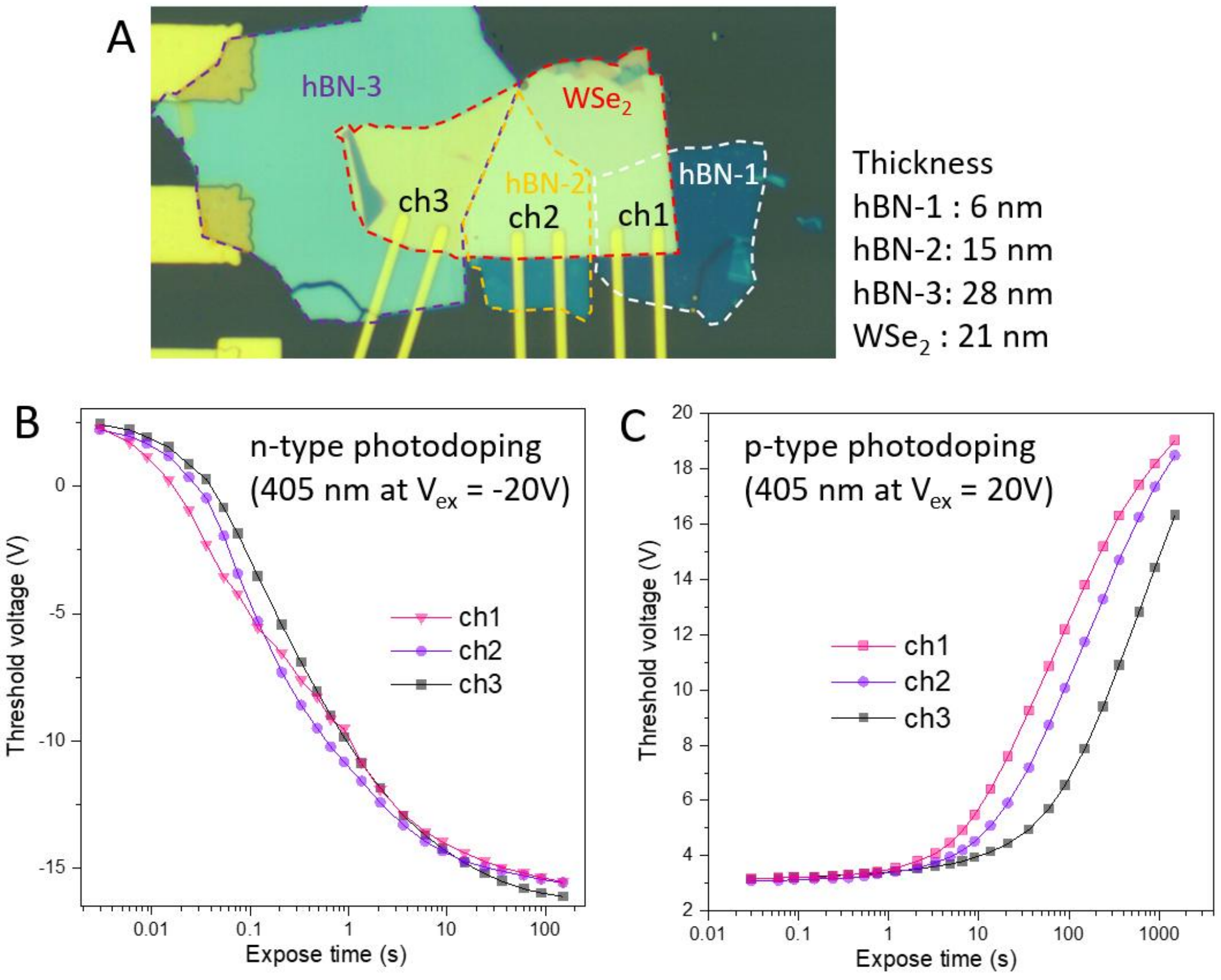

**Figure S4. Photodoping efficiencies on devices with different hBN thicknesses.** A) Optical microscope image of the $WSe_2$/hBN heterostructure field-effect transistors with different bottom hBN thicknesses of 6 nm (hBN-1), 15 nm (hBN-2), and 28 nm (hBN-3), but the thickness of $WSe_2$ channel is the same and 21 nm. Thus, the three channels (ch1, ch2, and ch3) have varied bottom hBN thickness, while other parameters such as the channel width, length, and $WSe_2$ thickness are almost the same. The p-type Si substrate with the $SiO_2$ thickness of 90 nm was used for this device. B) The shift of the threshold gate voltage according to the n-type photodoping obtained by exposing 405 nm laser with a power of 17 μW at $V_{ex}$ = -20 V for varied exposure durations, which are compared between the channels with different bottom hBN thicknesses. C) The same with B), but for the case of p-type photodoping with the 405 nm exposure at $V_{ex}$ = 20 V.

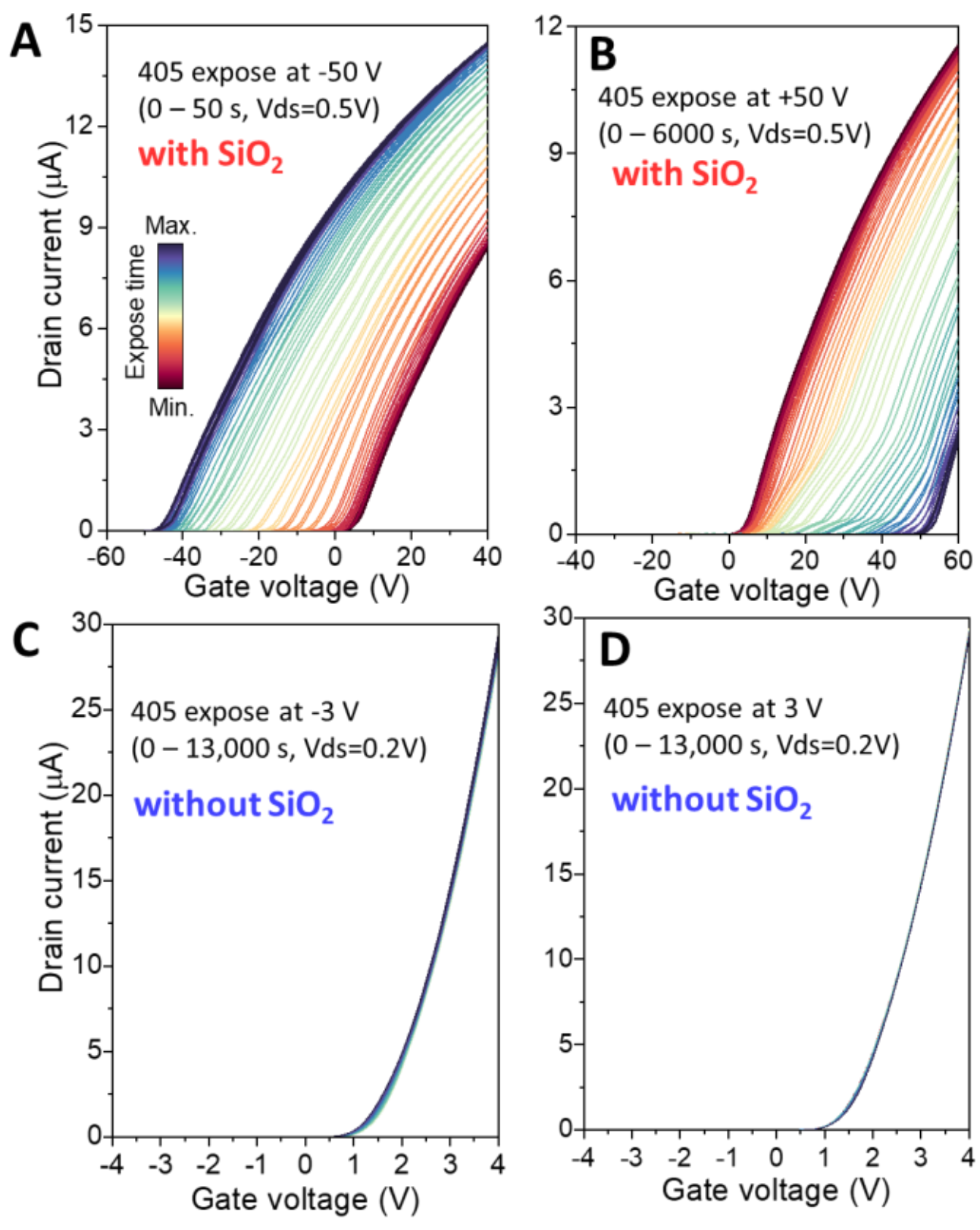

**Figure S5. Transfer-curve shifts as a function of 405 nm exposure time compared between the FETs with and without $SiO_2$ layer.** A) Exposure duration dependent transfer curve evolution of a $WSe_2$ FET on hBN/$SiO_2$/Si structure for the 405 nm exposure at $V_{Ex}$ = -50 V for n-type photodoping after the neutral condition is reached by the laser exposure at $V_{Ex}$ = 0 V. The device with SiO2 has 16 nm WSe2 and 38 nm bottom hBN, while the FET without SiO2 has 33 nm WSe2 and 35 nm thick bottom hBN. B) The same with A), but for the 405 nm exposure at $V_{Ex}$ = +50 V for p-type photodoping. C) Exposure duration dependent transfer curve evolution of a $WSe_2$ FET on hBN/Si structure for the 405 nm exposure at $V_{Ex}$ = -3 V for n-type photodoping after the neutral condition. D) The same with C), but for the 405 nm exposure at $V_{Ex}$ = +3 V for p-type photodoping. If the $SiO_2$ layer is removed the 405 nm laser exposure does not induce photodoping of $WSe_2$ for both n-type and p-type polarity.

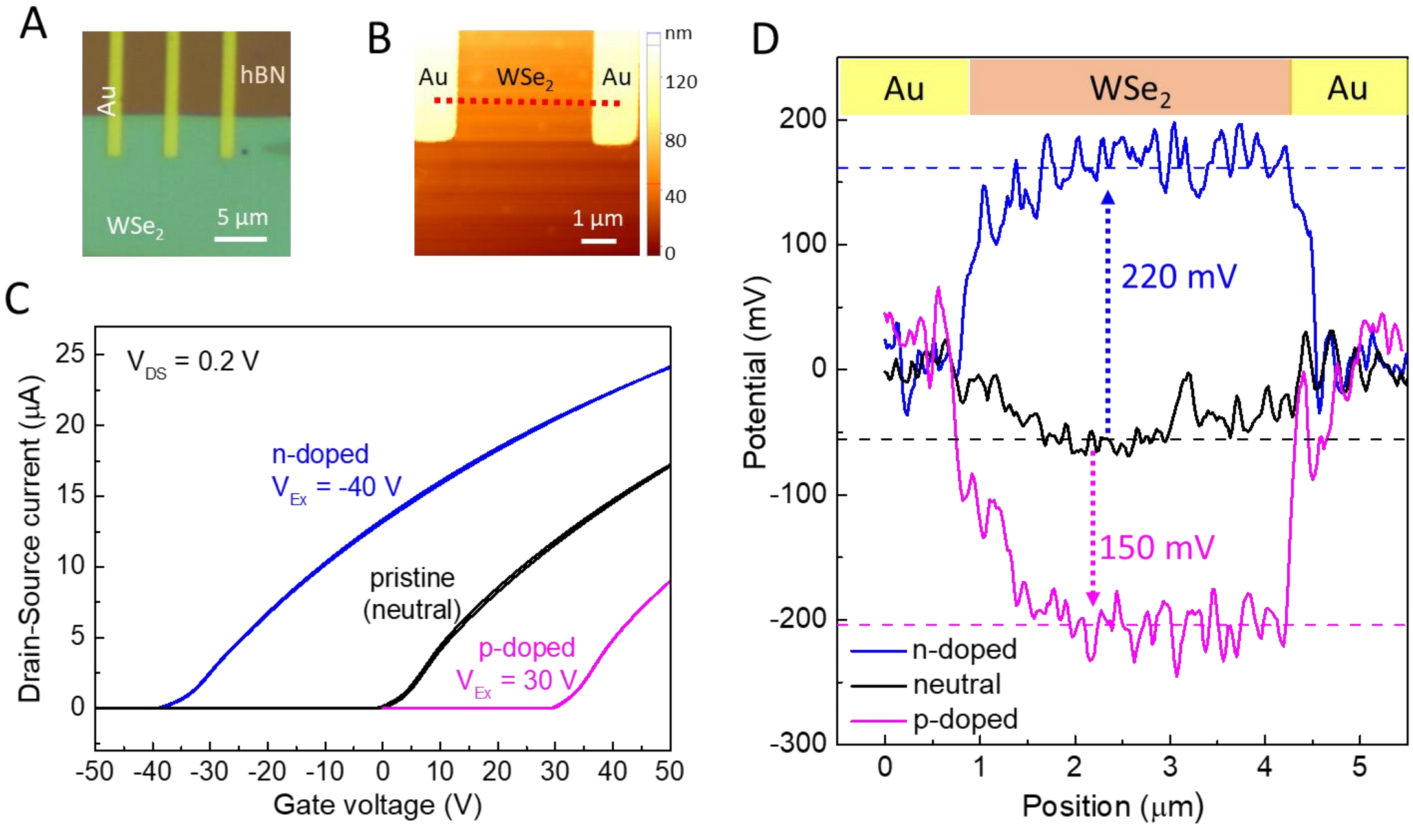

**Figure S6. Kelvin probe force microscopy to confirm charge transfer in $WSe_2$ after the n-type and p-type photodoping.** A) Optical microscopy image of a $WSe_2$ FET device with bottom hBN, but without top hBN for KPFM measurements. B) AFM topography image of the $WSe_2$ channel and the Au metal electrode. The thickness of $WSe_2$ (18 nm) and bottom hBN (38 nm) could be obtained from this image. C) Transfer curves obtained after n-type doping with $V_{Ex}$=-40 V, neutral doping with $V_{Ex}$=0 V, and p-type doping with $V_{Ex}$= 40 V. D) Contact potential difference profiles along the red line in the AFM topography for the three doping conditions, in which an offset was added to make the Au electrode potential zero as a reference. The $WSe_2$ potential shift of 220 mV after the n-type photodoping demonstrates the real electron carrier doping in the $WSe_2$ channel, while the negative shift by -150 mW after the p-type photodoping indicates the hole carrier generation. The surface topography and contact potential difference between the tip and surface were measured simultaneously with tip height kept constant at 15 nm.

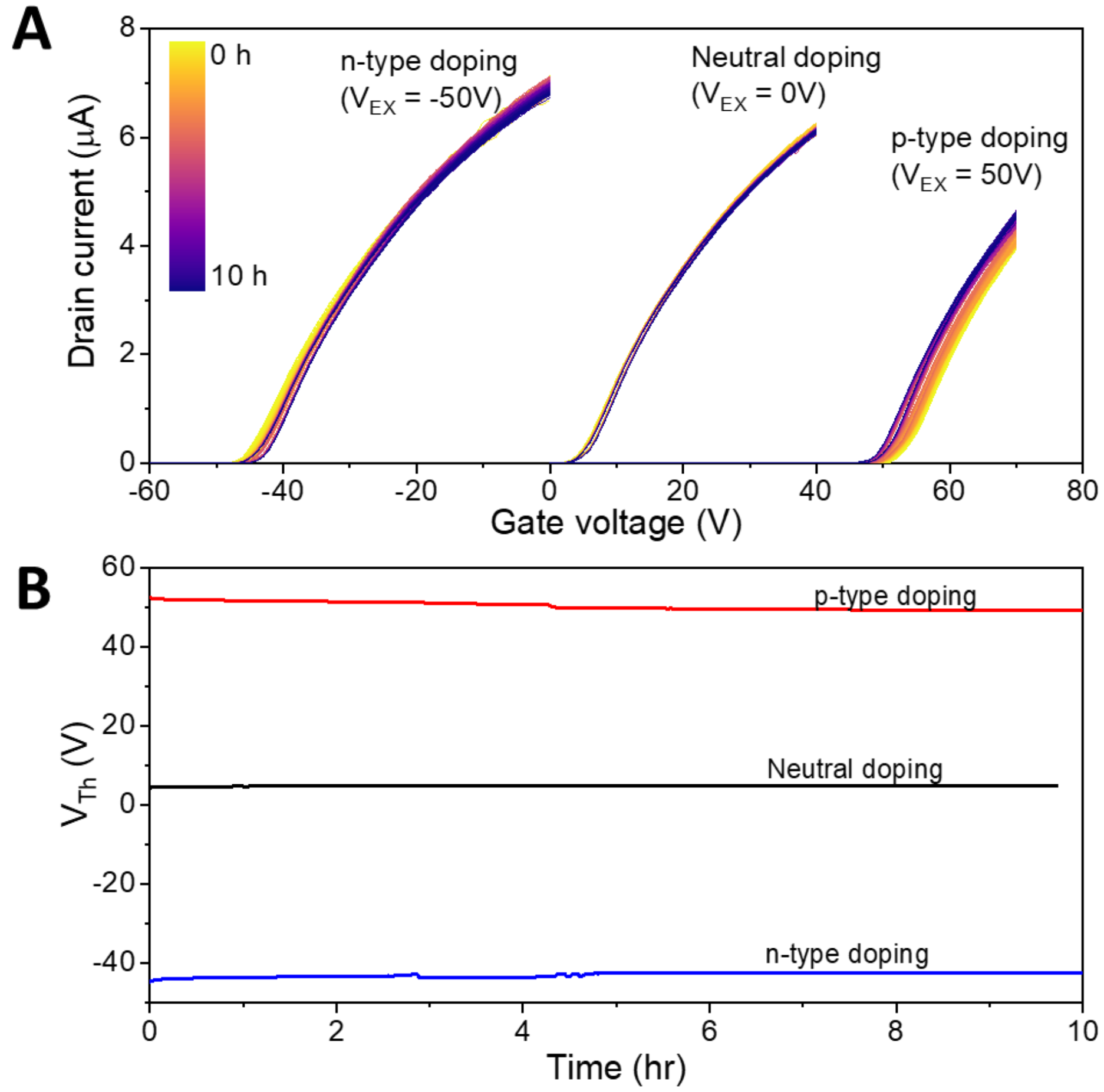

**Figure S7. Persistency of photodoping with time.** A) Drain current versus gate voltage curves (transfer curve) obtained at different elapse times up to 10 hours after the encapsulated $WSe_2$ channel was photodoped by exposing 405 nm lasers at $V_{Ex}$ = -50 V, 0 V, and + 50 V for the n-type, neutral, and p-type doping, respectively. The transfer curve was repeatedly acquired at an interval of 100 s util the time has passed by 10 hours. B) The drift of the threshold voltage with the elapsed time after the $WSe_2$ channel was treated for the p-type, neutral, and n-type photodoping. The threshold voltage of 52.5 V just after p-type photodoping ($V_{Ex}$= +50V) has changed to 49.3 V after 10 hours, corresponding to 6.4% decline. For the case of n-type photodoping ($V_{Ex}$= -50V), the doping concentration has reduced by 4.6% after 10 hours.

**First-principles calculation of hBN defects**

To investigate the possible contribution of hBN defect states to the photodoping process, we performed first-principles calculations for representative intrinsic defect configurations in bulk hBN. Vacancy-related and substitutional defects, including $V_N$, $V_B$, $N_i$, and $B_i$, were considered in order to evaluate their electronic structures and optical excitation energies.

The calculations were carried out within the framework of density-functional theory using the Vienna ab initio Simulation Package (VASP).[S3,S4] The projector-augmented wave (PAW) method was employed with a plane-wave cutoff of 400 eV. Exchange-correlation effects were treated using the screened-hybrid HSE06 functional[S5] together with Grimme's D3 dispersion correction with Becke-Johnson damping.[S6] Intrinsic point defects in bulk hBN were modeled in a 5 × 5 × 2 supercell, and all atomic positions were relaxed until the residual forces were below 0.03 eV $Å^{-1}$. VASP input files were generated using the VISE package, and defect calculations were managed with the pydefect code.[S7] Vertical excitation energies were corrected following the finite-size correction scheme proposed by Gake et al. for charged defects in periodic supercell calculations.[S8]

The calculated vertical excitation energies of these defect states span a wide range within the bandgap of hBN. In particular, several donor- and acceptor-like defect states are found to be optically accessible under 405 nm (3.06 eV) excitation, while some defect levels, such as VB and Ni, remain accessible even under lower-energy excitation corresponding to 640 nm (1.94 eV). These results indicate that optical activation of hBN defect states is, in principle, possible under both excitation conditions used in this study. However, despite the energetic accessibility of these defect states, the experimentally observed photodoping under 640 nm illumination is negligible, especially for p-type doping. This inconsistency suggests that defect excitation within hBN alone cannot account for the observed photodoping behavior. Combined with the experimental observations presented in the main text, these results support the conclusion that the dominant photodoping mechanism arises from charge transfer and trapping associated with defect states in $SiO_2$ at the hBN/$SiO_2$ interface, rather than from bulk defect states in hBN.

## Previous reports on the photodoping in 2D material/hBN optoelectronic devices

| Title | Journal | Year | Interpretation | Laser energy | Materials |
|---|---|---|---|---|---|
| [L1] Photoinduced doping in heterostructures of graphene and boron nitride | Nat. Nanotechnology | 2014 | Donor-like defect states in hBN | 2.15–2.6 eV | Graphene/BN /$SiO_2$/Si |
| [L2] Spatial Control of Laser-Induced Doping Profiles in Graphene on Hexagonal Boron Nitride | ACS Applied Materials and Interfaces | 2016 | Defects in hBN [L1], charges stopped at interface between hBN/$SiO_2$ | 2.33 eV (532 nm) | hBN-encap. graphene on $SiO_2$/Si substrate |
| [L3]Two-dimensional multibit optoelectronic memory with broadband spectrum distinction | Nat. Communications | 2018 | Donor-like defects states in hBN [L1] | 405, 638, 515, 473 nm | 1L$WSe_2$/BN/ $SiO_2$/Si |
| [L4] Nonvolatile and Programmable Photodoping in $MoTe_2$ for Photoresist-Free Complementary Electronic Devices | Advanced Materials | 2018 | Donor-like defect states in hBN [L1, L3] | 405 nm (3.06eV) | Multilayer $MoTe_2$/BN on $SiO_2$/Si & $MoTe_2$/BN/ Graphene on $SiO_2$/Si |
| [L5] Dynamically controllable polarity modulation of $MoTe_2$ field-effect transistors through ultraviolet light and electrostatic activation | Science Advances | 2019 | $MoTe_2$ excitation with tunneling hBN, defect states at the BN/SiO2 interface, Defect in hBN, | 254 nm | Multilayer $MoTe_2$/BN/ $SiO_2$/Si and $MoTe_2$/BN/ metal electrode |
| [L6] Reversible Photo-induced Doping in $WSe_2$ Field Effect Transistors | Nanoscale | 2019 | Defect states in hBN | 460, 640 nm, and laser with 1.8–2.6 eV | Multilayer $WSe_2$/BN/ $SiO_2$/Si |
| [L7] Photoinduced Doping To Enable Tunable and High-Performance Anti-Ambipolar $MoTe_2$/$MoS_2$ Heterotransistors | ACS Nano | 2019 | Defect states at the BN/$SiO_2$ interface, Defects in hBN [L1] | 254 nm | Multialyer $MoS_2$/hBN/Si O2/Si, Multialyer $MoTe_2$/BN/ $SiO_2$/Si |
| [L8] Semiconductor channel-mediated photodoping in h-BN encapsulated monolayer MoSe2 phototransistors | 2D Materials | 2019 | Electron-donor localized states are present in the hBN or at the MoSe2/hBN interface | 750–850 nm | hBN-encapsulated monolayer $MoSe_2$ on $SiO_2$/Si |
| [L9] Gate-tunable non-volatile photomemory effect in $MoS_2$ transistors | 2D Materials | 2019 | Trapped charges in gate-insulator interface (Graphite and hBN interface) | 488, 775 nm | Monolayer $MoS_2$/BN/ graphite |
| [L10] Two-dimensional electronic devices modulated by the activation of donor-like states in boron nitride | Nanoscale | 2020 | Donor-like states in BN [L1] | 220nm | Multilayer $WSe_2$/BN/met al electrode on the SiO2/Si |
| [L11] $WSe_2$ Homojunction p–n Diode Formed by Photoinduced Activation of Mid-Gap Defect States in Boron Nitride | ACS Applied Materials & Interfaces | 2020 | Mid-gap donor-like states (or defect states) in BN | 220 nm | $WSe_2$/BN/ $SiO_2$/Si |
| [L12] Nonvolatile van der Waals Heterostructure Phototransistor for Encrypted Optoelectronic Logic Circuit | ACS Nano | 2022 | Photogenerated carriers of $WSe_2$ and trapped at the BN/$Al_2O_3$ interface | 405 nm | Multilayer $WSe_2$/BN/ $Al_2O_3$ |
| [L13] Two-Dimensional Heterostructure Complementary Logic Enabled by Optical Writing | Small science | 2024 | Defect states in hBN | 355, 532 nm | Multilayer $WSe_2$/BN on $SiO_2$/Si |

| [L14] Optically Programmable Smart $WSe_2$/hBN Heterostructure Gas Sensors | ACS Applied Materials & Interfaces | 2025 | Defect states in hBN | UV | Multilayer W $Se_2$/hBN on $SiO_2$/Si |
|---|---|---|---|---|---|

Table S1: Reference list on photoinduced doping in 2D materials/hBN device

In Table S1, we list the previous reports on the photodoping in 2D material/hBN optoelectronic devices. Early studies attribute the phenomenon to donor-like defect states in hBN as possible origin [L1–L4]. Later reports suggest that other mechanisms such as carrier excitation in the channel [L5] and the hBN/gate(graphite) interface [L9]. Nevertheless, as shown in Table S1, many papers accept hBN defects as the primary cause. In our study, we could reveal photoexcitation of the defect in the $SiO_2$ region is the primary origin of photodoping, rather hBN defect states in our system. We expect that further systematic studies across different material and device architectures will help clarify the generality of this mechanism.

(S1) Blaga, C.; Labordet Alvarez, A.; Balgarkashi, A.; Banerjee, M.; Fontcuberta, I. M. A.; Dimitrievska, M. Unveiling the complex phonon nature and phonon cascades in 1L to 5L $WSe_2$ using multiwavelength excitation Raman scattering. Nanoscale Adv. 2024, 6 (18), 4591-4603. DOI: 10.1039/d4na00399c

(S2) Zhao, W.; Ghorannevis, Z.; Amara, K. K.; Pang, J. R.; Toh, M.; Zhang, X.; Kloc, C.; Tan, P. H.; Eda, G. Lattice dynamics in mono-and few-layer sheets of $WS_2$ and $WSe_2$. Nanoscale 2013, 5 (20), 9677-9683.

(S3) Kresse, G.; Furthmüller, J. Efficiency of Ab-initio Total Energy Calculations for Metals and Semiconductors Using a Plane-wave Basis Set. Comput. Mater. Sci. 1996, 6 (1), 15-50.

(S4) Kresse, G.; Joubert, D. From Ultrasoft Pseudopotentials to the Projector Augmented-wave Method. Phys. Rev. B 1999, 59 (3), 1758-1775.

(S5) Heyd, J.; Scuseria, G. E.; Ernzerhof, M. Hybrid Functionals Based on a Screened Coulomb Potential. J. Chem. Phys. 2003, 118 (18), 8207-8215.

(S6) Grimme, S.; Antony, J.; Ehrlich, S.; Krieg, H. A Consistent and Accurate Ab Initio Parametrization of Density Functional Dispersion Correction (DFT-D) for the 94 Elements H-Pu. J. Chem. Phys. 2010, 132 (15), 154104.

(S7) Kumagai, Y.; Tsunoda, N.; Takahashi, A.; Oba, F. Insights into Oxygen Vacancies from High-throughput First-principles Calculations. Phys. Rev. Mater. 2021, 5 (12), 123803.

(S8) Gake, T.; Kumagai, Y.; Freysoldt, C.; Oba, F. Finite-size Corrections for Defect-involving Vertical Transitions in Supercell Calculations. Phys. Rev. B 2020, 101 (2), 020102.